\documentclass[sigconf, nonacm]{acmart}

\usepackage{enumitem}
\setlist{noitemsep, topsep=0pt, parsep=0pt, partopsep=0pt}

\newcommand{\stitle}[1]{\vspace{1ex}\noindent\textbf{#1}}
\usepackage{xspace}
\usepackage{multirow}

\newcommand{\nikos}[1]{{\color{red} [nikos: #1]}}

\newcommand{\hide}[1]{}

\usepackage{enumitem}
\setlist{noitemsep, topsep=0pt, parsep=0pt, partopsep=0pt}

\usepackage{fancyvrb}
\usepackage[ruled, linesnumbered, vlined]{algorithm2e}
\usepackage[noend]{algorithmic}
\usepackage{subcaption}     % For subfigures
\usepackage{caption}

\SetKwInOut{Input}{Input}
\SetKwInOut{Output}{Output}
\SetKwInOut{Variables}{Variables}
\SetKwComment{comm}{\hfill$\triangleright$\ }{}

\usepackage{titlesec}

\titlespacing*{\section}
  {0pt}   % left margin
  {1.5ex} % space before
  {0.8ex} % space after
\titlespacing*{\subsection}
  {0pt}
  {1.2ex}
  {0.6ex}

\usepackage{tikz}
\usepackage{xcolor}

\usepackage{cleveref}

\usepackage[most]{tcolorbox}
\newcommand{\resquest}[1]{\begin{tcolorbox}[
    colback=gray!2,
    colframe=gray,
    boxrule=0.5pt,
    left=2pt,
    right=2pt,
    top=2pt,
    bottom=2pt,
    boxsep=1pt,
    before skip=4pt,
    after skip=4pt
]
{\em #1}\end{tcolorbox}}

\begin{document}

%\title{SIMD-powered Multidimensional Indexing}
\title{ANN Search: Recall What Matters}
% \\
%     \LARGE [Experiment, Analysis \& Benchmark]
%\subtitle{\huge \textbf{[Experiment, Analysis \& Benchmark]}}

%%
%% The "author" command and its associated commands are used to define the authors and their affiliations.
\author{Dimitris Dimitropoulos}
\orcid{0009-0005-1874-829X}
\affiliation{%
  \institution{{ University of Ioannina \& Archimedes, Athena RC}}
  \streetaddress{University of Ioannina}
  \city{Ioannina}
  \state{Greece}
}
\email{ddimitropoulos@cs.uoi.gr}

\author{Nikos Mamoulis}
\orcid{0000-0003-3423-4895}
\affiliation{%
  \institution{{ University of Ioannina \& Archimedes, Athena RC}}
  \streetaddress{University of Ioannina}
  \city{Ioannina}
  \state{Greece}
}
\email{nikos@cs.uoi.gr}

\begin{abstract}
Approximate Nearest Neighbor (ANN) search has become a core primitive in information retrieval and modern machine learning tasks, from classification to retrieval-augmented generation. The community evaluates and tunes ANN algorithms primarily on their throughput at a given Recall@k, the fraction of true exact neighbors retrieved. 
We argue that 
what really matters in ANN search is the quality of the retrieved results and not 
their overlap with the true kNN set.
We show that using Recall@k to assess retrieval quality forces unnecessary computational overhead and investigate replacing it by 1/Ratio@k, the inverse approximation ratio.
1/Ratio@k
evaluates the differences between the distances of the retrieved and true neighbors.
It is judge-free, hyperparameter-free, and computable from standard ANN benchmark inputs alone. We benchmark state-of-the-art ANN algorithms across diverse datasets spanning a wide range of intrinsic dimensionalities, evaluating the two metrics comprehensively across efficiency, downstream classification, and retrieval-augmented generation.  
On the efficiency axis, optimizing for 1/Ratio@k reaches operational quality thresholds at a substantially lower computational cost than Recall@k. In downstream tasks, performance indicators (label precision, semantic similarity, BERTScore, and LLM-graded quality) remain highly stable even when Recall@k drops significantly. The inverse approximation ratio, on the other hand, closely mirrors this stability, tracking true utility much better than Recall@k. Ultimately, while Recall@k overstates the true cost of approximation, 1/Ratio@k offers a more accurate, deployable proxy for actual ANN quality.
\end{abstract}

\maketitle

% %%% do not modify the following VLDB block %%
% %%% VLDB block start %%%
% \pagestyle{\vldbpagestyle}
% \begingroup\small\noindent\raggedright\textbf{PVLDB Reference Format:}\\
% \vldbauthors. \vldbtitle. PVLDB, \vldbvolume(\vldbissue): \vldbpages, \vldbyear.\\
% \href{https://doi.org/\vldbdoi}{doi:\vldbdoi}
% \endgroup
% \begingroup
% \renewcommand\thefootnote{}\footnote{\noindent
% This work is licensed under the Creative Commons BY-NC-ND 4.0 International License. Visit \url{https://creativecommons.org/licenses/by-nc-nd/4.0/} to view a copy of this license. For any use beyond those covered by this license, obtain permission by emailing \href{mailto:info@vldb.org}{info@vldb.org}. Copyright is held by the owner/author(s). Publication rights licensed to the VLDB Endowment. \\
% \raggedright Proceedings of the VLDB Endowment, Vol. \vldbvolume, No. \vldbissue\ %
% ISSN 2150-8097. \\
% \href{https://doi.org/\vldbdoi}{doi:\vldbdoi} \\
% }\addtocounter{footnote}{-1}\endgroup
% %%% VLDB block end %%%

% %%% do not modify the following VLDB block %%
% %%% VLDB block start %%%
% \ifdefempty{\vldbavailabilityurl}{}{
% \vspace{.3cm}
% \begingroup\small\noindent\raggedright\textbf{PVLDB Artifact Availability:}\\
% The source code, data, and/or other artifacts have been made available at \url{\vldbavailabilityurl}.
% \endgroup
% }
% %%% VLDB block end %%%

\section{Introduction}
Approximate Nearest Neighbor (ANN) search is a foundational operation in modern data management,
finding application in 
% . Vector databases built on ANN indices serve a growing share of 
information retrieval, recommender systems, and retrieval-augmented generation.
%workloads, and the algorithms that drive them are the subject of ongoing research across the database and machine learning communities. 
Because exact nearest neighbor search is computationally prohibitive at scale,  production systems trade accuracy for efficiency. The central empirical question in ANN research is therefore: {\em how much ANN query throughput  in Queries-Per-Second (QPS)
can we achieve at a given accuracy level?}
%how much exactness, traded for how much speed?

The community  answers this question largely under a specific definition of accuracy; Recall@$k$, defined as the fraction of the exact $k$NN set retrieved by A$k$NN search, has become the standard measure of retrieval quality in ANN evaluation. 
% Queries-per-second (QPS), which captures how many queries the system can serve in unit time is the de facto efficiency metric. 
Algorithms are typically compared via the QPS-vs-Recall curve, which plots throughput against retrieval quality across configurations. Established ANN benchmarks \cite{DBLP:journals/is/AumullerBF20, DBLP:journals/corr/abs-2505-17810, DBLP:journals/pacmmod/LiYLZCM25} rank algorithms by their throughput at fixed Recall thresholds, and state-of-the-art ANN algorithms \cite{DBLP:journals/pacmmod/GaoL24, DBLP:journals/corr/abs-2603-05180, DBLP:journals/pami/MalkovY20, DBLP:conf/icml/GuoSLGSCK20, DBLP:journals/pacmmod/WeiLLPP25, DBLP:journals/pacmmod/GaoL23, DBLP:journals/pacmmod/KuffoKB25, DBLP:conf/icde/PaparrizosELEF22, DBLP:journals/pacmmod/GaoGXYLW25, DBLP:journals/pacmmod/GouGXL25, DBLP:journals/pacmmod/WangZH25} are designed and tuned to optimize the QPS-vs-Recall@$k$ trade-off. The QPS-vs-Recall@$k$ curve has shaped more than a decade of ANN research.

In this paper, we
question the appropriateness of Recall as a target accuracy measure. We  
argue that the QPS-vs-Recall convention is misaligned with what ANN algorithms actually deliver, and that the community %has been paying a substantial computational cost to 
tunes ANN methods to optimize for a metric that overstates the difficulty of the task. We investigate whether \(1/\mathrm{Ratio}@k\),
%\nikos{it's better not to propose evaluation using ratio, but to investigate the effectiveness of ratio as an accuracy measure and the benefits that it brings when it replaces recall in ANN algorithms: high actual precision at higher QPS. We also want to investigate whether the relative performance of different ANN methods changes if we replace recall by ratio. Example: Is HNSW still the best method? Is the preprocessing cost still a bottleneck? Can PQ methods or collision methods be accurate enough while having a low preprocessing and memory cost?}
the inverse of the approximation ratio between the distances of an algorithm’s retrieved results and the true $k$NNs, is a more effective quality measure than Recall@\(k\). 
$1/\mathrm{Ratio}@k$ generalizes the
approximation ratio
definition
of Indyk and Motwani~\cite{DBLP:conf/stoc/IndykM98}
by averaging the per-position distance ratios across all $k$ positions.
Both Recall@\(k\) and \(1/\mathrm{Ratio}@k\) take values between $0$ and $1$
%\([0,1]\) 
with \(1\) indicating perfect retrieval, but they measure different things. Recall counts identifier matches, while \(1/\mathrm{Ratio}\) measures the quality of the retrieved results compared to the true $k$ NNs. 
%\stitle{The practicality 1/Ratio@k vs. Recall@k.}
%Early work on ANN \cite{DBLP:conf/stoc/IndykM98}
%used the approximation ratio to 
%measure retrieval quality 
%\nikos{is it exactly the same as the one we are using here?},
%but this was applied in medium-dimensionality data,
% with the approximation ratio  and in low-dimensional or low-difficulty settings, 
%where 
%Recall@$k$ and \(1/\mathrm{Ratio}\) 
%are aligned \nikos{who says so? do we have evidence for this?}. 
%Early work on ANN defined quality through the approximation ratio ~\cite{DBLP:conf/stoc/IndykM98}. The $c$-approximate nearest neighbor problem requires that a returned point $p'$ satisfies $\mathrm{dist}(q,p') \leq c \cdot \mathrm{dist}(q,p)$, where $c>1$. Our $1/\mathrm{Ratio}@k$ generalizes this definition 
%to the top-$k$ setting 
% by averaging the per-position distance ratios across all $k$ positions. 

\begin{figure}[htb]
    \centering
\includegraphics[width=0.9\columnwidth]{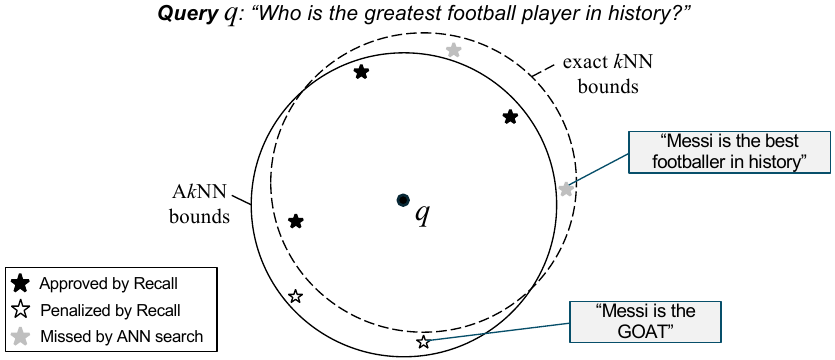}
    \vspace{-1mm}
    \caption{
        Approximate and exact $k$NN results in a RAG downstream task. The embeddings missed by the ANN algorithm are not necessarily of much higher quality compared to the retrieved ones not in the exact $k$NN set. Recall@$k$ gives unreasonably high penalty to retrieval quality.
    }
\label{fig:introex}
    % \vspace{-4mm} % Adjust spacing to fit tightly with the rest of the text
\end{figure}

\looseness=-1
When intrinsic dimensionality is low, successive neighbors are well-separated in distance, especially for small $k$ values, so 
%an algorithm that returns close points also returns the correct identifiers and 
$\mathrm{Recall}@k$ and $1/\mathrm{Ratio}@k$ are naturally aligned. As intrinsic dimensionality grows or $k$ becomes large, however, distances to successive neighbors concentrate~\cite{DBLP:conf/icdt/BeyerGRS99} and many near-equidistant candidates appear
% , making NN search harder~
\cite{DBLP:conf/sisap/0001C19}. As a result, an ANN algorithm can return points geometrically as close as the true neighbors but carrying different identifiers, causing $\mathrm{Recall}@k$ to degrade while $1/\mathrm{Ratio}@k$ remains high.
%This effect is illustrated in 
Figure \ref{fig:introex} exemplifies the exact and approximate $k$NN results for a RAG downstream task. Although Recall is low, retrieval quality is high. 
%, a divergence we quantify in Section~\ref{sec:exp_efficiency}.
This divergence is apparent in modern AI and ML applications, where object
embeddings are formed by hundreds to thousands of dimensions.
% , and at this scale the two metrics diverge, as we show in this paper. 
% \nikos{can we say that it is also more typical or beneficial for applications to use large $k$ values as opposed to $k=1$ or $k=10$?}
% This divergence 
Based on this, we conjecture that ANN methods may unnecessarily be tuned to achieve high recall levels, while ANN results of low recall may be of high quality, as reflected by $1/\mathrm{Ratio}@k$. 
% has measurable consequences in computational cost, 

\looseness=-1
\stitle{Other alternatives to Recall@$k$.}
Several recent works have questioned the effectiveness of Recall@\(k\) as a metric for ANN evaluation \cite{semantic, DBLP:journals/corr/abs-2512-12980, DBLP:journals/corr/abs-2507-00379}. Kuffo et al. \cite{semantic} introduce Semantic Recall and Tolerant Recall, both of which down-weight retrieval misses involving semantically irrelevant ground-truth neighbors. Semantic Recall requires an external LLM judge and access to the underlying raw documents; Tolerant Recall requires tuning a tolerance hyperparameter. Iceberg \cite{DBLP:journals/corr/abs-2512-12980} is a framework that re-ranks ANN algorithms by application-specific metrics such as Hit@\(\mathrm{K}\) for classification and retrieval tasks. 
% while this may not have a  significant effect in the accuracy of downstream tasks that employ ANN search. %remains unaffected.
%\nikos{what does \cite{DBLP:journals/corr/abs-2507-00379} propose?}
These efforts share our claim that Recall@\(k\) is misaligned with downstream utility, but their proposed solutions either have external dependencies or require hyperparameter tuning, which  limits their deployability. %Standard ANN benchmarks include only embeddings and ground-truth identifiers, not the raw documents or images those embeddings were derived from, and not application-task labels. 
%Semantic Recall cannot be computed without raw documents for the LLM judge to read; Iceberg's application metrics cannot be computed without task labels; and Tolerant Recall's threshold is data-dependent with no principled way to set it from embeddings alone. 
On the other hand, \(1/\mathrm{Ratio}@k\) requires only the embeddings indexed by the ANN algorithm; hence,  \(1/\mathrm{Ratio}@k\) is judge-free, hyperparameter-free, and computable from the data that standard ANN benchmarks already provide.

\stitle{Striving for high Recall@$k$ is expensive.}
To evaluate the appropriateness of \(1/\mathrm{Ratio}@k\) as opposed to Recall@$k$, 
across six datasets, we benchmark five state-of-the-art ANN algorithms: Annoy \cite{spotify_annoy}, a tree-based index; SuCo \cite{DBLP:journals/pacmmod/WeiLLPP25}, a collision-based framework; HNSW \cite{DBLP:journals/pami/MalkovY20}, the industry-standard graph-based index; RaBitQ \cite{DBLP:journals/pacmmod/GaoL24}, a quan\-tization-based method; and SymphonyQG \cite{DBLP:journals/pacmmod/GouGXL25}, a hybrid graph-and-quantization approach. We measure how much work each metric demands before reaching a target quality level (e.g., accuracy $\ge$ 0.95). We consider four cost factors: query time, distance computations per query, index build time, and memory footprint of the index. We find that for \(1/\mathrm{Ratio}@k\), it is substantially easier to reach operational thresholds compared to Recall@$k$ across all algorithms and datasets, with the gap growing with $k$.
Recall@$k$ and \(1/\mathrm{Ratio}\)@$k$ also disagree on which configurations meet the required quality standards, as Recall@$k$ largely underestimates quality. 
%To verify that 

\stitle{Achieving high Recall@$k$ is not crucial for downstream tasks.}
This disagreement matters in practice: configurations of ANN algorithms characterized as ``inadequate'' by not meeting high-Recall standards, actually deliver similar results as high-recall configurations, as Figure \ref{fig:introex} exemplifies.
We verify this across two downstream task families. In the first task, an image classification experiment on four benchmark datasets, Label Precision@100 barely changes as Recall drops to $0.4$, while \(1/\mathrm{Ratio}@100\) tracks it much more closely than Recall does.  Second, we apply a retrieval-augmented generation experiment 
on five benchmarks, measuring LLM answer quality by BERTScore F1, semantic similarity, and an LLM graded score.
% is preserved across the same recall range. 
Across all settings, the actual quality difference between high-recall and moderate-to-low recall retrieval is negligible, while 1/Ratio is shown to be a much better quality proxy.

\stitle{Summary of contributions}
%Our contributions can be summarised as follows:
\begin{itemize}
[noitemsep,topsep=2pt,parsep=0pt,partopsep=0pt,leftmargin=0.5cm]
\item We analyze the drawbacks of Recall@$k$ as a quality measure for ANN search and investigate the effectiveness of \(1/\mathrm{Ratio}@k\) as an appropriate replacement.
%quality measure for ANN search. 
%As to Recall@$k$ 
%    measure used in the majority of evaluation studies for ANN algorithms.
    % propose evaluating ANN algorithms by 1/Ratio@k, the inverse
    % approximation ratio, as an alternative quality metric to Recall@k.
    Unlike recently proposed alternatives to Recall@k \cite{semantic, DBLP:journals/corr/abs-2512-12980, DBLP:journals/corr/abs-2507-00379}, \(1/\mathrm{Ratio}@k\) is simultaneously judge-free, hyperparameter-free, and computable from the embeddings and ground-truth identifiers that standard ANN benchmarks already provide (see Table~\ref{tab:metric-comparison} for a comparison).
    
    \item We benchmark five ANN algorithms across 6 datasets, comparing the required cost to reach a quality target level under Recall@\(k\) and \(1/\mathrm{Ratio}@k\), in terms of query time, distance computations,  build time, and memory. Across all algorithms, reaching a Recall@$k$ target is substantially more expensive than reaching the same 1/Ratio@$k$ target, and the gap widens with $k$ (except for memory which is mainly unaffected).

    \item We show that Recall loss does not essentially degrade image classification quality. Under synthetic recall control on MNIST, Fashion-MNIST, CIFAR-10, and SVHN, Label Precision@100 is stable across Recall@100 \(\in [0.4,1.0]\). On the other hand, \(1/\mathrm{Ratio}@100\) tracks Label Precision substantially more closely than Recall does, providing evidence that \(1/\mathrm{Ratio}\) is a better predictor of downstream task quality.

    \item Similarly, we show that Recall loss does not degrade 
    retrieval-augmented generation quality. On three QA datasets evaluated 
    against human-annotated ground truth and two BEIR datasets evaluated 
    against an LLM perfect-recall baseline, answer quality measured by BERTScore F1, semantic similarity, and LLM graded score is preserved across the same recall range. \(1/\mathrm{Ratio}\) tracks this stability closely, confirming that it is a more accurate estimator of true downstream quality than Recall.
\end{itemize}

\begin{comment}
\begin{figure}[htb]
    \centering

    \includegraphics[width=\columnwidth]{figures/recall_vs_lp_figure.pdf}
    \vspace{-3mm}
    \caption{
        UNFINISHED: Recall vs label precision (LP) in ANNS. The query "7" has 5 exact nearest neighbors (3 sevens, 2 ones). The retrieved set matches only 2 of them (Recall@5 = 0.4) yet preserves the same LP@5 = 0.6.
    }
    \label{fig:lp_recall}
    \vspace{-4mm} % Adjust spacing to fit tightly with the rest of the text
\end{figure}
\end{comment}
\section{Related Work} \label{sec:relatedwork}
\subsection{ANN Algorithms}
ANN search has been studied for decades, and the methods that dominate fall into five broad families.
%possible four with hash. may need a paragraph for them as well i guess. we will see. 

\stitle{Tree-based} methods partition the vector space recursively along axis-aligned or random hyperplanes and route a query down the resulting hierarchy. Early instances such as KD-trees \cite{DBLP:journals/cacm/Bentley75} were designed for exact nearest neighbor search; approximate variants emerged by relaxing exactness and limiting the number of partitions visited during traversal \cite{10.1145/293347.293348}. %Later methods such as 
Annoy \cite{spotify_annoy} %adopted approximation as a first-class design goal, 
builds forests of independently randomized trees and merges their candidate sets. 
Tree-based methods 
underperform 
%do not scale with dimensionality, as their 
% While tree-based methods are simple to implement and support incremental updates, they do not scale with data dimensionality.
% are only su and dominate low-dimensional nearest-neighbor search. 
in embedding spaces, where dimensionality is high, as their partitions become ineffective: the volume of each leaf cell concentrates near its boundary, and search ends up traversing a large fraction of the tree 
%to identify the  nearest neighbors, eroding the speedup partitioning is meant to provide 
\cite{DBLP:conf/vldb/WeberSB98, DBLP:conf/stoc/DasguptaF08}.

\stitle{Hash-based} methods map vectors to short binary codes so that nearby points collide under the same hash bucket with high probability. Locality-Sensitive Hashing (LSH) \cite{DBLP:conf/stoc/IndykM98} provides sublinear query time with provable approximation guarantees by hashing 
to 
%queries into 
multiple independently constructed tables. LSH is applied in Euclidean spaces via p-stable distributions \cite{DBLP:conf/compgeom/DatarIIM04} and indexing  
with data-adaptive encoding schemes \cite{DBLP:journals/pvldb/WeiPLP24}. Learning-to-hash methods train hash functions tailored to the input dataset \cite{DBLP:journals/pami/WangZSSS18}. Recently, collision-based frameworks such as SuCo \cite{DBLP:journals/pacmmod/WeiLLPP25} generalize the hashing paradigm by counting subspace collisions as a proxy for Euclidean distance, while retaining theoretical guarantees on result quality. 
Recent benchmarks \cite{DBLP:journals/corr/abs-2505-17810, DBLP:journals/is/AumullerBF20, DBLP:journals/tkde/LiZSWLZL20}
show that 
%In practice, however, both 
hash-based and collision-based methods are outperformed in throughput by graph-based and quantization-based indices.

\stitle{Quantization-based} methods compress each vector into a short code and estimate distances directly on the codes.
% , avoiding the cost of holding and scanning full-precision vectors. 
Product quantization (PQ) \cite{DBLP:journals/pami/JegouDS11} splits a vector into sub-vectors and quantizes each sub-vector against a learned codebook; OPQ \cite{DBLP:journals/pami/GeHK014} adds a rotation that aligns the data with the PQ partition; ScaNN \cite{DBLP:conf/icml/GuoSLGSCK20} learns a quantizer that minimizes the inner-product estimation error. %specific to MIPS. 
Inverted-file (IVF) structures are typically layered on top to first restrict the search to a small number of clusters and then estimate distances on the codes inside those clusters. RaBitQ \cite{DBLP:journals/pacmmod/GaoL24}, a randomized 1-bit quantizer with a theoretical error bound, is the current state-of-the-art in this family. 
%A common limitation of 
Quantization-based methods depend on how well the learned codebooks or cluster partitions capture the data distribution: when the data is poorly fit by the quantizer, distance estimation errors can be high. 
% Gao and Long [15] show that PQ incurs more than 50\% average relative error on the MSong dataset, causing recall to remain below 60\% even with re-ranking. 
% More broadly, these methods require an offline training phase over the dataset to learn codebooks, which adds to the index construction cost and makes them sensitive to distributional assumptions \cite{DBLP:journals/tkde/XuTZ18}.

\stitle{Graph-based} methods precompute a navigable proximity graph over the dataset and answer queries by greedy traversal: starting from an entry point, the search moves to the neighbor closest to the query and repeats until no closer neighbor is found. Variants differ in how the graph is constructed. NSW \cite{DBLP:journals/is/MalkovPLK14} grows the graph incrementally with small-world properties; NSG \cite{DBLP:journals/pvldb/FuXWC19} sparsifies it for fewer hops; SSG \cite{DBLP:journals/pami/FuWC22} refines NSG with a satellite-system construction for improved robustness; and Vamana/DiskANN \cite{NEURIPS2019_09853c7f} adapts the construction to disk-resident indices. Graph-based methods currently dominate ANN benchmarks and production systems \cite{DBLP:journals/pacmmod/WangWKGZZ25, DBLP:journals/pvldb/WangXY021}; HNSW \cite{DBLP:journals/pami/MalkovY20}, the hierarchical variant of NSW, is the most established index in this family and is the default vector index in PostgreSQL, Pinecone, Milvus \cite{Milvus}, and Elasticsearch \cite{DBLP:journals/pacmmod/WangWKGZZ25}.
While being very efficient, graph-based methods 
are limited by their large 
memory footprint \cite{DBLP:conf/icde/YueX0TL24, DBLP:journals/corr/abs-2510-25401} and high 
construction cost 
% have well-known practical limitations. Their memory footprint is large, as every vector must reside in memory together with its adjacency list, making billion-scale indices difficult to fit on a commodity machine \cite{DBLP:journals/corr/abs-2510-25401}. Index construction is also expensive: building the graph requires computing distances between each inserted vector and a large candidate set, and increasing graph quality (via larger values for parameters $M$ or $efConstruction$) amplifies this cost substantially 
\cite{DBLP:journals/pacmmod/WangWKGZZ25, DBLP:journals/pvldb/WangXY021}.

\stitle{Hybrid Methods} combine techniques from different families; e.g., a graph for navigation and quantized codes for distance computations. NGT-QG \cite{ngtqg} and SymphonyQG \cite{DBLP:journals/pacmmod/GouGXL25} are representative of this hybrid line, inheriting the routing efficiency of graph traversal and the throughput of quantized distance estimation.

\subsection{Benchmarking conventions}
The research community evaluates ANN indexes almost exclusively along Recall@k versus queries-per-second (QPS) curves. This convention was established by early ANN benchmarks \cite{DBLP:journals/is/AumullerBF20} and remains the reference framework for the field. The recently released VIBE benchmark \cite{DBLP:journals/corr/abs-2505-17810}, which 
tests
% includes very high dimensional embeddings and tests 
out-of-distribution workloads
on very high dimensional embeddings, still uses Recall as the primary quality metric.
% modernizes this pipeline with embedding-era and out-of-distribution workloads, but still uses Recall as the primary quality metric. 
DARTH \cite{DBLP:journals/pacmmod/ChatzakisPP25} turns Recall into a service-level objective via adaptive early termination at user-declared Recall targets.
%, a direct evidence that achieving a fixed Recall threshold is the operational goal the community optimises for. 
In general, optimizing for Recall is the main objective 
%in The same pattern holds 
of published work on ANN indexing during the past several years
% state-of-the-art (SOTA) from the last years: a long line of recent indexes 
\cite{DBLP:journals/pacmmod/GaoL24, DBLP:journals/corr/abs-2603-05180, DBLP:journals/pami/MalkovY20, DBLP:conf/icml/GuoSLGSCK20, DBLP:journals/pacmmod/WeiLLPP25, DBLP:journals/pacmmod/GaoL23, DBLP:journals/pacmmod/KuffoKB25, DBLP:conf/icde/PaparrizosELEF22, DBLP:journals/pacmmod/GaoGXYLW25, DBLP:journals/pacmmod/GouGXL25, DBLP:journals/pacmmod/WangZH25, DBLP:journals/pvldb/ChenZHJW24, DBLP:journals/corr/abs-2405-19504}, where headline results are reported by Recall-QPS curves and ANN indices tune build and search hyperparameters to hit pre-specified Recall thresholds. Recall@k is therefore not merely one evaluation choice among many: it is the optimization target that has shaped both the design of ANN search methods and modern retrieval systems.
%in the embedding era.

\subsection{Criticism on Recall}
Few recent studies question whether Recall@k measures what practitioners actually need.
%maybe we need more info here
Kuffo et al. \cite{semantic} argue that traditional Recall penalizes algorithms for missing ground-truth neighbors that are actually not semantically relevant to the query.
%, calling this the problem of retrieving ``the mathematically closest noise'' rather than meaningful answers. 
% This is the inverse of what we criticize: optimising for recall marks as irrelevant neighbors that are marginally outside the exact $k$NN set. 
They propose two replacements: \emph{Semantic Recall}, the fraction of semantically relevant ground-truth neighbors an algorithm retrieves, where relevance is determined by human judges or by LLMs and \emph{Tolerant Recall}, which counts a retrieved item as a match whenever its distance to the query is close enough to the true neighbor's distance, with closeness set by a tunable tolerance. Neither replacement is simultaneously judge-free and hyperparameter-free. Semantic Recall requires ground truth
% an external judge (human or LLM) and access to the source content the embeddings were derived from (text passages, images, etc.), 
which standard ANN benchmarks do not provide. 
Tolerant Recall 
is conceptually similar to 1/Ratio@k, but 
requires tuning a 
% avoids the judge but introduces a 
tolerance hyperparameter 
% that has to be re-tuned 
per dataset. 
% Tolerant Recall is conceptually the closest alternative to 1/Ratio@k, as both assess whether retrieved vectors are geometrically close to the true neighbors rather than requiring strict identifier matches. However, Tolerant Recall introduces a tolerance threshold that must be calibrated per dataset and their proposed heuristic for setting it is approximate and acknowledged as requiring further refinement.

Iceberg \cite{DBLP:journals/corr/abs-2512-12980} evaluates SOTA vector-similarity-search methods in downstream tasks, scoring them by application-level metrics such as label recall, hit rate, and matching score. 
It is  shown that
% When algorithms are ranked by these task-level metrics rather than by Recall-QPS curves, the ranking changes substantially: 
an algorithm that wins on Recall is often not the best algorithm based on the task-level metric (e.g. label recall). 
However,
this methodology  is expensive to apply: each evaluation needs the source data the embeddings come from, a separate embedding model, and task-specific labels, none of which standard ANN benchmarks provide.

Wang et al. \cite{DBLP:journals/corr/abs-2507-00379} criticize  Recall-based benchmarking from a different angle. Standard ANN benchmarks report a single number, the mean of Recall@k across the query set, which can hide a long tail of hard queries: two systems with the same average Recall can behave very differently across queries, one delivering consistent quality, and the other excelling on most queries while performing poorly on a minority of hard queries. They suggest counting quality, rather than averaging by proposing Robustness-$\delta$@K, which reports the fraction of queries whose Recall exceeds a user-chosen floor $\delta$. They validate this metric in an end-to-end RAG case study where systems with similar mean Recall produce noticeably different answer quality, and conclude that Recall ranks systems inconsistently, 
%with itself 
depending on whether one averages or counts. 
Like 1/Ratio, Robustness-$\delta$@k is judge-free and requires only benchmark-standard inputs;
however (i) it is still based on Recall, so it disregards the similarity between retrieved and exact results, and (ii) 
it relies on a user-chosen floor parameter $\delta$ that must be tuned per dataset, limiting its deployability.

%\stitle{\bf Main Contribution.}
Our paper joins this line of skepticism against Recall@k, showing that it systematically overstates the cost of approximation in distance computations, query time, and index build time. We also show that the configurations Recall@$k$ finds inadequate in fact deliver equivalent results in downstream tasks. 
Going one step further compared to previous work, we compare Recall@$k$ against %To make this argument empirically, we need a second metric to compare Recall against, and we propose
1$/\mathrm{Ratio}$@$k$, which,
unlike alternative measures \cite{semantic,DBLP:journals/corr/abs-2512-12980,DBLP:journals/corr/abs-2507-00379}, is
a judge-free, hyperparameter-free quality measure computable from the same inputs ANN benchmarks already provide; see  Table~\ref{tab:metric-comparison}. We experimentally show that 1$/\mathrm{Ratio}$@$k$ (i) converges to its operational thresholds substantially faster than Recall@$k$ and (ii) matches the quality of downstream tasks more closely than Recall@$k$. 

\begin{comment}
\nikos{No need to discuss specific results here} (Figures X and Z), and it tracks downstream task quality far more tightly: on the image classification benchmark, sweeping configurations across Recall $\in [0.4, 1.0]$ keeps $1/\mathrm{Ratio}@k$ within a $1\text{-}3.5\%$ band, and downstream task quality, measured for example by Label Precision@100, stays comparably stable across the same sweep (Figure Z). On the RAG benchmarks, across three BEIR datasets evaluated against a perfect-recall LLM baseline and three QA datasets evaluated against human-annotated ground truth, both sweeping $\mathrm{Recall}@10  \in [0.4, 1.0]$, BERTScore~F1 and LLM-graded scores vary by under 
$X\%$ across the approximate retrieval range $[0.4, 0.9]$, a range 
over which Recall itself swings 50\% and $1/\mathrm{Ratio}$ varies by around 3 percentage points, showing that $1/\mathrm{Ratio}$ is a far closer proxy for answer quality than Recall is.

\end{comment}

\begin{table}
\centering
\footnotesize
\setlength{\tabcolsep}{3pt}
\caption{Proposed alternatives to Recall@$k$}
\label{tab:metric-comparison}
\resizebox{\columnwidth}{!}{%
\begin{tabular}{lccc}
\toprule
\textbf{Metric} & \textbf{Judge-free} & \textbf{Hyperparameter-free} & \textbf{Benchmark-inputs-only} \\
\midrule
% Recall@$k$ &
% \checkmark &
% \checkmark &
% \checkmark \\

Semantic Recall \cite{semantic} &
$\times$ &
\checkmark &
$\times$ \\

Tolerant Recall \cite{semantic} &
\checkmark &
$\times$ &
\checkmark \\
Iceberg task metrics \cite{DBLP:journals/corr/abs-2512-12980} &
\checkmark &
\checkmark &
$\times$ \\

Robustness-$\delta$@K \cite{DBLP:journals/corr/abs-2507-00379} &
\checkmark &
$\times$ &
\checkmark \\

\textbf{$1/\mathrm{Ratio}@k$ (our proposal)}&
\checkmark &
\checkmark &
\checkmark \\
\bottomrule
\end{tabular}%
}
\end{table}

\section{Preliminaries and Compared ANN Methods}
\label{sec:preliminaries}

\subsection{Problem definition}
\label{sec:prelim-setup}

Let $X$ be a set of $N$ vectors in $\mathbb{R}^D$. Depending on the application, nearest neighbor search aims to find the most relevant vectors to a query by either minimizing a distance function $dist(\cdot,\cdot)$ or maximizing a similarity function $sim(\cdot,\cdot)$. We use Euclidean distance ($L_2$) for the cost-accuracy and classification experiments, which has conventionally been used in the evaluation of state-of-the-art algorithms 
\cite{DBLP:journals/pacmmod/GouGXL25, DBLP:journals/pacmmod/GaoL24, DBLP:journals/pami/MalkovY20}
and in 
ANN benchmarking
\cite{DBLP:journals/pami/WangZSSS18, DBLP:journals/tkde/LiZSWLZL20}.
For the RAG experiments, we use cosine similarity, as the inner product of $L_2$-normalized vectors, matching the standard configuration for dense text retrieval \cite{DBLP:conf/emnlp/KarpukhinOMLWEC20}.
Since we can set $dist(x,y) = 1 - sim(x,y)$, a unified definition of similarity search using 
$dist(\cdot,\cdot)$ is as follows. 
% To unify these two paradigms and evaluate retrieval quality consistently across the rest of this paper, we formalise similarity-based search as distance minimisation by defining the effective distance as $dist(x,y) = 1 - sim(x,y)$. Consequently, 
For a query $q \in \mathbb{R}^D$ and an integer $k \geq 1$, the \emph{exact $k$} nearest neighbors of $q$ form the set $N_k(q) \subseteq X$ of the $k$ vectors in $X$ closest to $q$ under $dist(\cdot,\cdot)$. \emph{Approximate nearest neighbor (ANN)} search finds a set $\widetilde{N}_k(q) \subseteq X$ of size $k$ that approximates $N_k(q)$.

\subsection{Accuracy Measures for ANN Search}
\label{sec:prelim-metrics}

% \subsection{Recall@$k$ and 1/Ratio@k}
% \label{sec:prelim-metrics}

We evaluate ANN algorithms with two quality metrics
%, both taking values 
between $0$ and $1$, with $1$ indicating perfect retrieval
w.r.t.
%, with respect 
the exact $k$NN set.

\stitle{Recall@$k$} is the fraction of retrieved true $k$ neighbors:
\vspace{-1mm}
\begin{equation}
\mathrm{Recall}@k(q) = \frac{\lvert \widetilde{N}_k(q) \cap N_k(q) \rvert}{k}.
\end{equation}
Recall@$k$ is a set-overlap measure: it counts identifier matches between the returned set $\widetilde{N}_k(q)$ and the true $k$NN set $N_k(q)$, and ignores how close the missed candidates are to the true neighbors.

\stitle{1/Ratio@k} measures how close the returned neighbors are to the true $k$NN set. Let $d_i(q)$ denote the distance from $q$ to the $i$-th true neighbor by distance and $\widetilde{d}_i(q)$ the 
distance to the $i$-th returned neighbor by distance.
% , with both lists sorted in ascending order of distance. 
Following Indyk and Motwani's original ratio formulation \cite{DBLP:conf/stoc/IndykM98}, we first define the approximation Ratio@$k$ as the mean of per-position distance ratios:
\vspace{-1mm}
\begin{equation}
    \text{Ratio}@k(q) = \frac{1}{k}\sum_{i=1}^{k} 
    \frac{\widetilde{d}_i(q)}{d_i(q)}.
\end{equation}
Since the true $k$NNs are by definition the closest points to $q$, each per-position ratio $\widetilde{d}_i(q) / d_i(q)$ lies in $[1, \infty)$, so $\text{Ratio}@k \geq 1$, with equality at perfect retrieval. We evaluate algorithms by its inverse:
\vspace{-1mm}
\begin{equation}
    \frac{1}{\text{Ratio}@k}(q) = \frac{k}{\sum_{i=1}^{k} 
    \frac{\widetilde{d}_i(q)}{d_i(q)}},
\end{equation}
which takes values in $(0, 1]$. A value of $1$ means the returned 
neighbors have the same distances to the query as the true neighbors, while a value 
below $1$ quantifies how much further out the returned set lies.
Recently, ANN evaluation has moved away from distance-based definitions of approximation quality, adopting Recall@k as a quality measure \cite{DBLP:journals/pacmmod/GaoL24, DBLP:journals/corr/abs-2603-05180, DBLP:journals/pami/MalkovY20, DBLP:conf/icml/GuoSLGSCK20, DBLP:journals/pacmmod/WeiLLPP25, DBLP:journals/pacmmod/GaoL23, DBLP:journals/pacmmod/KuffoKB25, DBLP:conf/icde/PaparrizosELEF22, DBLP:journals/pacmmod/GaoGXYLW25, DBLP:journals/pacmmod/GouGXL25, DBLP:journals/pacmmod/WangZH25, DBLP:journals/pvldb/ChenZHJW24, DBLP:journals/corr/abs-2405-19504}.

%\nikos{this paragraph is unnecessary; what you say here is a conclusion from the experiments; we have talked about this before and we will talk about this later. this section is about definitions and not about conclusions or results.}
%In low-dimensional or low-difficulty settings the two metrics agree, but
%in modern embedding spaces with hundreds to thousands of dimensions they
%can diverge substantially. The divergence is amplified by k: as k grows,
%the exact top-k set extends into a region of the distance distribution
%where many vectors are nearly equidistant from the query, so the
%probability that an ANN algorithm returns a geometrically close vector
%that carries a different identifier increases, penalising Recall while
%leaving 1/Ratio essentially unchanged. 

\subsection{Downstream Task Evaluation Metrics}\label{sec:downstreameval}

% \stitle{Label Precision at $k$ (LP@$k$)},
% used in the classification experiments (Section~\ref{sec:classification}),
% averages the fraction of the $k$ approximate NNs returned for each $q$ in a query set $Q$ that share the $q$'s class label:
% \begin{equation}
%     \text{LP@}k = \frac{1}{|Q|} \sum_{q\in Q}
%     \frac{\left|\{i \in \widetilde{N}_k(q) : \ell_i = \ell_q\}\right|}{k},
% \end{equation}
% where $\widetilde{N}_k(q)$ is the set of $k$ approximate neighbours returned
% for query $q$ and $\ell_q$ is its ground-truth class label.
% A value of 1.0 means every retrieved neighbour shares the $q$'s label for all $q\in Q$.

\stitle{Label Precision at $k$ (LP@$k$)},
used in the classification experiments (Section~\ref{sec:classification}),
% ,
% Label Precision at $k$
%It 
is the fraction of the $k$ approximate NNs returned for a query $q$ that share $q$'s class label:
\begin{equation}
    \text{LP@}k(q) = 
    \frac{\left|\{i \in \widetilde{N}_k(q) : \ell_i = \ell_q\}\right|}{k},
\end{equation}
where $\widetilde{N}_k(q)$ is the set of $k$ approximate neighbors returned
for query $q$ and $\ell_q$ is its ground-truth class label.
A value of 1.0 means every retrieved neighbor shares $q$'s label.

\stitle{Semantic Similarity} \cite{Milvus} is used in the RAG experiments (Section~\ref{sec:rag}). It
is the cosine similarity between dense sentence embeddings of the generated
answer and the reference answer, produced by \texttt{all-mpnet-base-v2}. A score of 1.0 indicates perfect semantic agreement with the reference.
%; lower values reflect increasing divergence.

\stitle{BERTScore F1} \cite{Zhang2019BERTScoreET} is used in the RAG experiments (Section~\ref{sec:rag}).
% BERTScore~\cite{Zhang2019BERTScoreET} 
It evaluates textual similarity between a candidate and a reference by matching, via cosine similarity, their contextual token embeddings (\texttt{distilbert-base-uncased}). Unlike Semantic
Similarity, which embeds the answer as a single vector, BERTScore operates at the token level. A score of 1.0 indicates perfect token-level alignment with the reference; lower values reflect increasing textual divergence.

%\textbf{LLM-graded score.}
%An LLM (Llama~3.1:8b) is prompted to score each generated answer on a
%1--10 scale given the question and the reference answer, providing a
%holistic quality judgment that complements the two automatic metrics above.

\subsection{Local intrinsic dimensionality}
\label{sec:prelim-lid}
In accordance to previous studies \cite{DBLP:conf/sisap/0001C19} and also based on our experimental results, performance of ANN search 
% we assert that $k$NN search complexity 
is more closely related to the dataset's intrinsic difficulty than to the ambient dimensionality $D$.
%, though no single measure fully explains cross-dataset variation.
% which is not directly  related to how hard nearest-neighbour search actually is. 
\emph{Local Intrinsic Dimensionality} (LID) estimates the effective dimensionality of the data; higher LID has been shown to correlate with increased query difficulty in ANN search \cite{DBLP:conf/sisap/0001C19, DBLP:journals/tkde/LiZSWLZL20}.
%manifold in a small neighbourhood, and which %
In our evaluation we compute the LID for each dataset under its own distance function, using the TwoNN estimator \cite{DBLP:journals/corr/abs-1803-06992}, which derives LID from the ratio of distances to the first and second nearest neighbors and is robust to the choice of neighborhood size. LID values in our benchmark range from 7.19 (PubMedQA) to 32.35 (Gist), covering a wide range of difficulty levels. Throughout the paper, we examine how the divergence between Recall@$k$ and $1/\mathrm{Ratio}@k$ relates to LID across datasets and algorithms.

\subsection{ANN algorithms benchmarked}
\label{sec:prelim-algorithms}

In our efficiency experiments,
we benchmark five algorithms spanning the graph, quantization, hybrid, tree, and collision-based categories, serving as prominent representatives of their respective families.
% in high-dimensional spaces. 
% ANNOY~\cite{spotify_annoy}, a tree-based index used in Spotify; SuCo~\cite{DBLP:journals/pacmmod/WeiLLPP25}, an advanced collision-based framework, HNSW~\cite{DBLP:journals/pami/MalkovY20}, considered the industry standard for graph-based methods; RaBitQ~\cite{DBLP:journals/pacmmod/GaoL24}, the current state-of-the-art quantiser; and SymphonyQG~\cite{DBLP:journals/pacmmod/GouGXL25}, the current state-of-the-art graph-quantisation hybrid approach.
%All five are benchmarked in the efficiency experiments.
Our classification and RAG experiments use HNSW only, as it is the established production default \cite{DBLP:journals/pacmmod/WangWKGZZ25}.

\stitle{Annoy} \cite{spotify_annoy} is a tree-based index built from a forest of random projection trees. During construction, each tree recursively partitions the space by selecting two points at random and splitting along the hyperplane equidistant between them, continuing until leaf nodes hold only a few vectors. At query time, the algorithm traverses all trees using a priority queue, preferring branches closer to the query and exploring both sides of a split when the query falls near the boundary. Candidates from the visited leaves are collected and ranked. Two parameters control the speed–accuracy trade-off: the number of trees built ($\mathit{num\_trees}$) and the number of nodes inspected during search ($\mathit{search\_k}$).

\stitle{SuCo} \cite{DBLP:journals/pacmmod/WeiLLPP25} is a collision-based framework that estimates vector similarity by counting subspace collisions. The index splits the vector space into several disjoint, lower-dimensional subspaces and clusters each using an Inverted Multi-Index (IMI) \cite{DBLP:conf/cvpr/BabenkoL12}, forming a joint codebook of centroid combinations. At query time, the query vector is projected into each subspace to identify its nearest cells. Any database vector sharing an activated cell registers a collision. Vectors that collide in enough subspaces are forwarded for re-ranking by exact Euclidean distance. The speed–accuracy trade-off is controlled by the subspace layout, the collision threshold, and the candidate limit for re-ranking.

\looseness=-1
\stitle{HNSW} \cite{DBLP:journals/pami/MalkovY20} is a hierarchical proximity graph. Vectors are inserted incrementally, and each vector is assigned a maximum layer level drawn from a geometric distribution.
%, so that the number of nested layers grows logarithmically with $n$. 
Each layer is a navigable-small-world graph, with higher layers containing fewer points and longer-range edges. Search begins at the top layer with a single entry point and proceeds greedily: at each layer, the algorithm maintains a candidate list of size $\mathit{efSearch}$ (the search beam) and descends to the next layer once a local minimum is reached. Index construction is controlled by two parameters, $M$ (the per-element degree budget) and $\mathit{efConstruction}$ (the build-time candidate-list size). We follow the parameter naming used by the Faiss implementation \cite{DBLP:journals/tbd/JohnsonDJ21}.

\stitle{RaBitQ} \cite{DBLP:journals/pacmmod/GaoL24} is a randomized 1-bit quantization method. Vectors are first normalized onto the unit hypersphere and then rotated via a random orthogonal transformation. Each coordinate of a rotated vector is then mapped to a single bit, giving a $D$-bit code per vector. From these bit codes, RaBitQ constructs an unbiased estimator of inner products and Euclidean distances with a theoretical error bound that scales with $1/\sqrt{D}$, i.e., the bound tightens as the (embedded) dimensionality grows. Distance estimation is implemented with bitwise operations or SIMD-based routines for high throughput.

\stitle{SymphonyQG} \cite{DBLP:journals/pacmmod/GouGXL25} integrates RaBitQ-style quantization with a navigable-graph routing structure. Each vertex stores the quantization codes of its graph neighbors in a packed layout that can be scanned sequentially with FastScan, a SIMD-based batched distance estimation routine. The standard quantize-then-rerank pipeline used in earlier graph--quantization hybrids is avoided: SymphonyQG searches directly on quantized distances and uses a refinement strategy based on multiple estimated distances to compensate for the accuracy loss this would otherwise introduce.
\section{Research Questions (RQs)}

To evaluate the appropriateness of Recall@$k$ in assessing ANN search performance and the 
utility of our suggested $1/\mathrm{Ratio}@k$ measure, we structure our study around four research questions.

\stitle{RQ1. Cost of optimizing for Recall@$k$ vs. 1/Ratio@$k$.}
% ANN evaluation has converged on a single metric, Recall@k, to define what it means for a configuration to be ``good enough.'' Hence, 
Index construction parameters, search budget parameters, and quantization parameters are all conventionally tuned to achieve a target Recall@$k$ threshold in ANN search, and benchmarks rank algorithms by their QPS at fixed retrieval quality levels.  
As Recall@$k$ and 1/Ratio@$k$
% , the standard quality measure, only considers overlap with the exact $k$NN set, disregarding how close the returned neighbours are to the query in distance, and in high-dimensional embedding spaces these two notions 
can diverge substantially, we first quantify the  overhead incurred to satisfy Recall@$k$ targets when distance quality (measured by $1/\mathrm{Ratio}$@$k$) is already satisfied.

\resquest{
%\begin{quote}
\textbf{RQ1.} How much  overhead does optimizing ANN indices for Recall@k impose, in query time, distance computations, index build time, and memory footprint, compared to optimizing for 1/Ratio@k?}
%\end{quote}

\stitle{RQ2. Stability of algorithm rankings.}
% Recall-QPS trade-offs form the standard baseline for evaluating, comparing, and selecting ANN algorithms. 
As replacing Recall by a different quality metric
may change the ANN algorithm that performs best under a target QPS throughput or construction cost budget, 
a natural question is whether the conclusions by Recall-based experimental analyses generalize. 
%then the community's current conception about the best performing ANN search algorithms would be challenged.
%rankings reflect the metric as much as the algorithm. 
We therefore investigate whether replacing Recall with $1/\mathrm{Ratio}$ reorders algorithm rankings.
\resquest
{
\textbf{RQ2.} Does replacing Recall@$k$ with $1/\mathrm{Ratio}$@$k$ change the relative ranking of ANN algorithms under fixed budgets of query throughput, distance computations, index build time, or memory?
}

\begin{comment}

[Pending: When we present the Pillar-1 we answer this via the Ratio-QPS]

Under Recall@100 $\ge 0.95$: HNSW cannot reach this threshold under any parameter setting, meaning it is unreachable and ranks last (behind RaBitQ).
Under $1/\mathrm{Ratio}$@100 $\ge 0.95$: HNSW is fully reachable and achieves $5,338$ QPS, outperforming RaBitQ ($2,989$ QPS).
This demonstrates that changing the quality metric directly reorders algorithm rankings (swapping HNSW and RaBitQ).

\end{comment}

\stitle{RQ3. Geometric explanation.}
If there is a divergence between the achieved $1/\mathrm{Ratio}$ and Recall by an index configuration, the gap should be predictable from properties of the data rather than from algorithmic artifacts.
LID is the most established data property in the ANN literature, and neighborhood size $k$ is a natural search parameter: as $k$ grows, the query boundary enclosing the true nearest neighbors expands, increasing the probability that an ANN algorithm returns a geometrically close vector with a different identifier, penalizing Recall while leaving $1/\mathrm{Ratio}$ largely unchanged. We therefore investigate whether the gap between Recall and $1/\mathrm{Ratio}$ degradation grows with LID and with $k$, and whether this connection holds across algorithm families.

\resquest
{
\textbf{RQ3.} Is the divergence between Recall@k and $1/\mathrm{Ratio}@k$ systematically related to the intrinsic dimensionality of the dataset's embedding space and the size of the retrieved neighborhood $k$?}

\stitle{RQ4. Downstream task quality at ``inadequate'' Recall}.
Recall@$k$ labels a configuration as inadequate if it fails to return a target percentage of the $k$ most relevant objects. The implicit assumption is that objects missing from the exact $k$NN set 
%identifiers \nikos{what does `identifier' mean in this sentence? Do you mean `missing objects'?} 
%carry information that
crucially affect the performance of downstream tasks. We evaluate this assumption directly across two task families. In an image classification task, we measure whether Label Precision degrades when retrieval is performed at lower Recall levels. In a retrieval-augmented generation task, we measure whether LLM answer quality, judged by both automatic metrics and human-annotated gold answers, degrades when retrieval Recall drops. If neither task degrades materially, the configurations Recall dismisses are not inadequate in an operational sense, since the retrieved objects not in the exact $k$NN set can still assist the downstream task to achieve satisfactory accuracy.

\resquest
{
\textbf{RQ4.} Do ANN index configurations that have low Recall@k deliver significantly degraded downstream task quality compared to high-Recall configurations?
% , or do they deliver equivalent quality at lower cost? 
Furthermore, between Recall@k and 1/Ratio@k, which metric serves as a more faithful proxy for true real-world performance of downstream tasks?
}

\section{Experimental Setup}\label{sec:setup}

%\stitle{Setup.}
% We conduct three sets of experiments: an analysis of cost-accuracy trade-offs across algorithms (RQ1--RQ3), an evaluation of the impact of Recall and 1/Ratio on image classification accuracy (RQ4), and an evaluation of their impact on retrieval-augmented generation (RAG) quality (RQ4).
% Our evaluation is conducted on a machine with an 11th Gen Intel\textsuperscript{\textregistered} Core\texttrademark{} i7-11700K
% (3.60\,GHz), 32\,GB RAM, and AVX-512 support. All indexing and query code is C++ compiled with \verb|gcc| 11.4 on Ubuntu 22.04 LTS using each project's official configuration. We utilise OpenMP and native threading policies for multi-threading across all indexing and query phases. The implementations of   Annoy\footnote{\url{https://github.com/spotify/annoy}}, SuCo\footnote{\url{https://github.com/WeiJiuQi/SuCo}}, FAISS-HNSW\footnote{\url{https://github.com/facebookresearch/faiss}}, and SymphonyQG\footnote{\url{https://github.com/gouyt13/SymphonyQG}} fully utilise AVX-512 instructions (either via explicit intrinsics or compiler auto-vectorisation). In contrast, RaBitQ is evaluated using AVX2, as its codebase\footnote{\url{https://github.com/gaoj0017/RaBitQ}} lacks support for AVX-512.

Our evaluation is conducted on a machine with an 11th Gen Intel\textsuperscript{\textregistered} Core\texttrademark{} i7-11700K
(3.60\,GHz), 32\,GB RAM, and AVX-512 support. All indexing and query code is C++ compiled with \verb|gcc| 11.4 on Ubuntu 22.04 LTS using each project's official configuration. We utilize OpenMP and native threading policies for multi-threading across all indexing and query phases. The implementations of   Annoy (\url{github.com/spotify/annoy}), SuCo (\url{github.com/WeiJiuQi/SuCo}), FAISS-HNSW (\url{github.com/facebookresearch/faiss}), and SymphonyQG (\url{https://github.com/gouyt13/SymphonyQG}) fully utilize AVX-512 instructions (either via explicit intrinsics or compiler auto-vectorization). In contrast, RaBitQ (\url{https://github.com/gaoj0017/RaBitQ}) is evaluated using AVX2, as it lacks support for AVX-512.

\stitle{Algorithms and Parameter Sweeps.}
We compare Annoy~\cite{spotify_annoy}, SuCo~\cite{DBLP:journals/pacmmod/WeiLLPP25}, 
HNSW~\cite{DBLP:journals/pami/MalkovY20} (FAISS implementation \cite{DBLP:journals/tbd/JohnsonDJ21}), RaBitQ \cite{DBLP:journals/pacmmod/GaoL24}, and SymphonyQG~\cite{DBLP:journals/pacmmod/GouGXL25}, sweeping each method's primary hyperparameters to meet our target quality thresholds.
{\textbf{Annoy}:} number of random projection trees %$\text{num\_trees} 
$\in \{10, 20, 50, 100\}$, search boundary limits %$\text{search\_k} 
$\in \{100, 500,$ $1000, 5000\}$.
{\textbf{SuCo}:} collision ratios $\in \{0.02, 0.03, ..., 0.06\}$, candidate ratios $\in \{0.001, 0.002, ..., 0.006\}$; we test multiple subspace partition configurations, starting from a baseline tuned to roughly 16 dimensions per subspace, with the total number of subspaces scaled by variations of $/3$, $/2$, $\times 2$, and $\times 3$.
{\textbf{RaBitQ}:} cluster sizes $C \in \{32, 64, ... , 4096\}$ in powers
of two, $N_{\text{probe}}$ in powers of two from $2$ to $256$ with
$N_{\text{probe}} \leq C$, and a fast k-means (2 iterations) versus
standard 20-iterations based training.
{\textbf{HNSW}:} $M \in \{16, 32, 48, 64\}$,
$\mathit{efConstruction} \in \{32, 64,$ $128\}$,
$\mathit{efSearch} \in \{32, 64, 128, 256\}$.
{\textbf{SymphonyQG}:} $M \in \{32, 64\}$ (aligned to the 32-bit SIMD
boundary), $\mathit{efConstruction} \in \{32, 64,$ $128, 256, 512\}$,
$\mathit{efSearch} \in \{32, 64, 128, 256, 512, 1024\}$; quantization
training fixed at $3$ iterations (codebase default).
%quantisation-training regime.

\stitle{Cost-Accuracy Datasets and Metrics.}
The cost-accuracy trade-off experiments (RQ1--RQ3) use six $D$-dimensional datasets of varying intrinsic dimensionalities (LID), modalities, and cardinalities (Table~\ref{tab:datasets}).
$L_2$ stands for Euclidean distance (the smaller the better) and IP for (normalized) inner product (the larger the better).
We sweep $k \in \{1, 20, 50, 100\}$ and, for each configuration, record
Recall@$k$, $1/\text{Ratio}@k$, QPS, index build time, and distance
computations per query, building one Pareto frontier per quality metric.

\begin{table}[h]
    \caption{Dataset Characteristics and LID}
    \label{tab:datasets}
    \vspace{-0.05in}
    \centering
    \resizebox{\linewidth}{!}{%
    \begin{tabular}{@{}l l c c c c c c@{}}
        \toprule
        \textbf{Dataset} & \textbf{Type} & \textbf{Card. ($N$)} & \textbf{$D$} & \textbf{\#Queries ($Q$)} & \textbf{$k$} & \textbf{Metric} & \textbf{LID} \\
        
        \midrule
        \multicolumn{8}{l}{\textbf{Cost-Accuracy Experiments}} \\
        \midrule
        \textbf{Gist \cite{gist-dataset}}                  
            & Image & 1,000,000 & 960  & 1,000  & 100 & $L_2$ & 32.35 \\
            
        \textbf{SimpleWiki \cite{datasets}}            
            & Text & 260,372   & 3,072 & 1,000  & 100 & $L_2$ (norm.) & 27.45 \\
            
        \textbf{ImageNet \cite{datasets}}          
            & Image & 1,281,167 & 640  & 1,000  & 100 & $L_2$ (norm.) & 17.85 \\
            
        \textbf{AGNews \cite{datasets}}                
            & Text & 769,382 & 1,024  & 1,000    & 100 & $L_2$ & 24.09 \\
            
        \midrule
        \multicolumn{8}{l}{\textbf{Cost-Accuracy and Classification Experiments}} \\
        \midrule
        \textbf{MNIST \cite{datasets, lecun2010mnist}}                 
            & Image & 60,000    & 784  & 10,000 & 100 & $L_2$ & 12.41 \\
            
        \textbf{Fashion-MNIST \cite{datasets, DBLP:journals/corr/abs-1708-07747}}         
            & Image & 60,000    & 784  & 10,000 & 100 & $L_2$ & 13.99 \\
            
        \midrule
        \multicolumn{8}{l}{\textbf{Classification Experiments}} \\
        \midrule
             
        \textbf{CIFAR-10 \cite{Krizhevsky2009LearningML}} 
            & Image & 50,000    & 3,072 & 10,000 & 100 & $L_2$ & 27.25 \\
            
        \textbf{SVHN \cite{netzer2011reading}} 
            & Image & 73,257    & 3,072 & 10,000 & 100 & $L_2$ & 18.99 \\
            
        \midrule
        \multicolumn{8}{l}{\textbf{Retrieval-Augmented Generation (RAG) Experiments}} \\
        \midrule
        \textbf{SciFact \cite{DBLP:conf/nips/Thakur0RSG21}}               
            & Text & 5,183     & 384  & 292    & 10  & IP (norm.) & 13.67 \\
            
        \textbf{NFCorpus \cite{DBLP:conf/nips/Thakur0RSG21}}              
            & Text & 3,633     & 384  & 117    & 10  & IP (norm.) & 10.58 \\
            
        \textbf{HotpotQA \cite{yang2018hotpotqa}}              
            & Text & 66,581    & 384  & 200    & 10  & IP (norm.) & 22.68 \\
            
        \textbf{MS-MARCO \cite{DBLP:conf/nips/NguyenRSGTMD16}}              
            & Text & 78,675    & 384  & 200    & 10  & IP (norm.) & 7.56 \\
            
        \textbf{PubMedQA \cite{DBLP:conf/emnlp/JinDLCL19}}              
            & Text & 3,358     & 384  & 200    & 10  & IP (norm.) & 7.19 \\
            
        \bottomrule
    \end{tabular}
    }
\end{table}

\stitle{Classification Datasets and Metrics.}
We evaluate the recall-ratio divergence on a downstream image-classification
task over MNIST, Fashion-MNIST, CIFAR-10, and SVHN. All four datasets use 10,000 test images as queries. MNIST, Fashion-MNIST and CIFAR-10 use the standard image training split as the base corpus and the complete test split as queries. For SVHN, we use the training split as the base corpus and the first 10,000 images of the test split as queries, matching the query count of the other three datasets. All benchmarks reuse the standard datasets and class labels from \url{huggingface.co/datasets/}.
Following recent works that isolate the effect of retrieval quality on downstream performance by constructing retrieval sets from controlled mixtures of gold and near-miss items \cite{DBLP:journals/corr/abs-2411-07396, DBLP:conf/emnlp/LiO25}, we evaluate performance using artificially synthesized retrieval sets at controlled recall levels $r \in \{0.4, 0.5, ..., 1.0\}$. This synthesis is necessary as HNSW typically achieves near-1.0 Recall, making a wide accuracy spectrum otherwise difficult to 
observe empirically. Specifically, for a target recall $r$, we construct a 100-item result set by mixing $100r$ exact top-100 items (the "gold set") with $100(1-r)$ items drawn in rank order from the nearest non-gold neighbors immediately beyond the gold set.

We report Label Precision at $k=100$ (LP@$100$), i.e., the fraction of returned
vectors sharing the query's class. We normalize the LP@$100$ of ANN results to the LP of the exact $k$NN set as follows. 
If $\mathrm{LP}@100(r)$ denotes LP at $k=100$ on the synthesized set at recall level $r$, the normalized LP@100 value is $\mathrm{LP}_{\mathrm{norm}}(r)=\frac{\mathrm{LP}@100(r)}{\mathrm{LP}@100(1.0)}$.
This is done because
% normalisation is needed because 
exact-search LP@100 varies widely across datasets (from 0.228 on CIFAR-10 to 0.887 on MNIST); dividing by the LP value at $r = 1.0$ removes these baseline differences and places LP, 1/Ratio, and Recall on a common scale where 1.0 denotes exact-search quality, isolating quality changes as recall drops. 
%If $\mathrm{LP}@100(r)$ denotes LP at $k=100$ on the synthesised set at recall level $r$, the normalised LP value is $\mathrm{LP}_{\mathrm{norm}}(r)=\frac{\mathrm{LP}@100(r)}{\mathrm{LP}@100(1.0)}$.

\stitle{RAG Datasets and Metrics.}
Documents and queries across  five RAG datasets are encoded  using ~\texttt{all-MiniLM-L6-v2} to $384$-dim vectors, which were $L_2$-normalized, and indexed with \texttt{hnswlib}. For HotpotQA, MS-MARCO, and PubMedQA, the retrieval corpus is the union of the candidate passages bundled with each query in the dataset, deduplicated where applicable; the corpus sizes in Table~\ref{tab:datasets} reflect this, not the full document collection.
We mirror the controlled evaluation protocol used in our classification setup, fixing the generation neighborhood context size to $k=10$. To construct the artificially synthesized retrieval sets for our target recall levels $r \in \{0.4, 0.5, ..., 1.0\}$, we replace $k(1-r)$ gold items with the nearest non-gold neighbors drawn in rank order from positions $k+1$ onward in the exact HNSW ranking.
For each target recall level $r \in \{0.4, 0.5, ..., 1.0\}$, we feed the 
mixed retrieval context into Llama~3.1:8b (temperature 0, via Ollama). 
To ensure a comprehensive evaluation, we analyze these datasets under two 
complementary reference paradigms. The first group, HotpotQA, MS-MARCO and PubMedQA (we use the first $200$ queries of the standard split) evaluates performance against independent, human-written ground-truth annotations, cross-validating the trends under a setup free from LLM self-evaluation bias. The second group, SciFact and NFCorpus (from the BEIR benchmark \cite{DBLP:conf/nips/Thakur0RSG21}) utilizes LLM-generated reference answers at $r=1.0$ (temperature 0.3) to isolate how approximate context alters the model's internal consistency. Each query is subject to a per-query generation timeout of 120 minutes; queries exceeding this limit are excluded from results. For SciFact, 292 of the 300 standard BEIR test queries produced valid results. For NFCorpus, 117 of the 323 BEIR test queries produced valid results; the remainder either had no relevance judgments or exceeded the timeout.
We evaluate downstream answer quality via Semantic Similarity 
%\cite{Milvus} 
(cosine similarity 
%over \texttt{all-mpnet-base-v2} embeddings), a 1-10 LLM grade, and a BERTScore F1 (\texttt{distilbert-base-uncased}). All quality metrics, along with $1/\text{Ratio}$  are normalised to the exact-search baseline ($r=1.0$). normalisation is necessary because raw quality scores have different scales and dataset-specific baselines. Even at perfect recall ($r = 1.0$), the LLM does not achieve a perfect score against independent ground truth (e.g. Semantic Similarity ranges from $0.517$ on MS-MARCO to $0.755$ on PubMedQA). normalising to $r = 1.0$ removes these baseline differences, isolating our primary quantity of interest: how much quality changes as recall drops.
over \texttt{all-mpnet-base-v2} embeddings), a 1-10 LLM grade, and a BERTScore F1 (\texttt{distilbert-base-uncased}), discussed in Section \ref{sec:downstreameval}. As in the classification setup, all quality metrics are normalized to the exact-search value ($r=1.0$). Even at $r=1.0$, absolute scores vary substantially across datasets (e.g. Semantic Similarity is 0.517 on MS-MARCO vs. 
0.755 on PubMedQA). normalizing to this value isolates 
%our quantity of interest: 
how much each metric changes as recall drops.

\section{Experimental Results and Analysis}\label{sec:exp}
\subsection{Cost-Accuracy Tradeoff (RQ1, RQ2, RQ3)} \label{sec:exp_efficiency}

We begin by quantifying the computational overhead that optimizing for
Recall@$k$ imposes over optimizing for 1/Ratio@$k$. For each algorithm and dataset, we sweep the full hyperparameter grid described in
Section~\ref{sec:setup}, recording QPS, distance
computations per query, index build time and memory footprint for every configuration at different Recall@$k$ and 1/Ratio@$k$ levels.
We report results at $k = 100$ by default, 
%throughout this subsection 
and analyze the effect of~$k$ at the end.

\begin{figure*}[t]
  \centering
  \includegraphics[width=\textwidth]{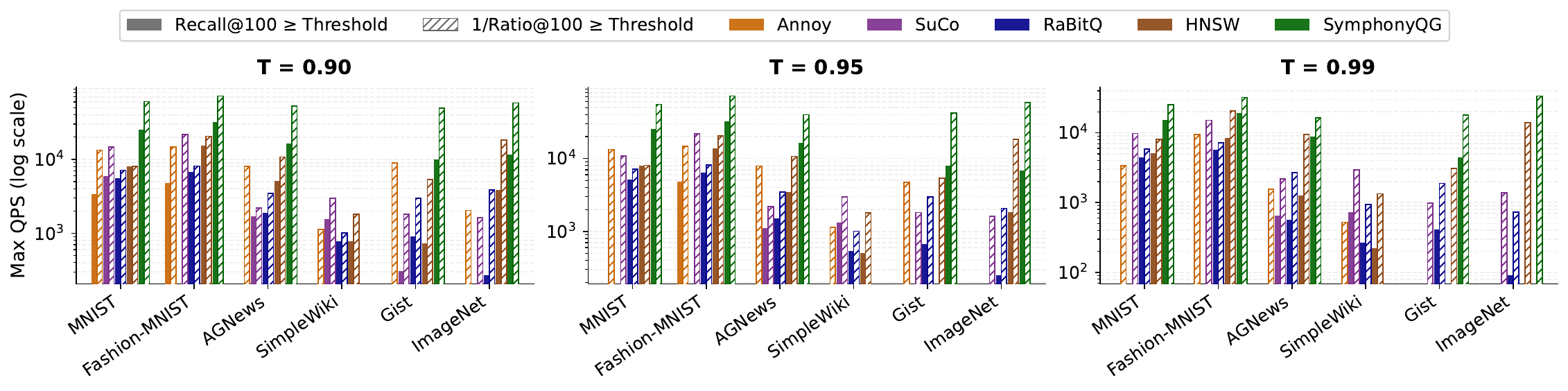}
  \caption{Maximum achievable QPS under Recall@100 and 1/Ratio@100 across target quality thresholds $T \in \{0.90, 0.95, 0.99\}$. Missing bars indicate the threshold was unreachable by any tested configuration.}
  \label{fig:qps_bars}
  \vspace{-2mm}
\end{figure*}

\stitle{QPS-quality trade-offs (RQ1, RQ3).}
Figure~\ref{fig:qps_bars} plots the maximum achievable throughput for configurations satisfying target quality thresholds under both metrics. Across all tested algorithms and datasets, the maximum throughput that can be achieved by a configuration that satisfies a given 1/Ratio threshold $T$ (e.g., at least $T=0.9$) is strictly higher than the maximum throughput of a configuration satisfying the same threshold $T$ when Recall is used. The gap between the solid and hatched bars represents the throughput that the community currently loses by tuning ANN algorithms to reach minimum Recall targets as opposed to 1/Ratio targets.
%as opposed to the corresponding 1/Ratio targets.

On low-LID datasets (MNIST, Fashion-MNIST), Recall and 1/Ratio have moderate QPS difference
% between the QPS performance of the two quality metrics is moderate 
for HNSW, RaBitQ, and SymphonyQG, reaching up to $2.25\times$. In such data, the true neighbors are well separated from the rest, so geometric closeness and identifier match are aligned. 
% SuCo and Annoy do not follow the same pattern even at low LID: 
SuCo cannot reach Recall@100~$\geq 0.95$ on either dataset, and Annoy fails on MNIST while exhibiting a $3.07\times$ speedup on Fashion-MNIST. On the other hand, SuCo and Annoy  satisfy 1/Ratio@100~$\geq 0.95$ at thousands of QPS.
% , indicating that tree-based and subspace-partition indices bear a high Recall penalty 
% % identifier-mismatch penalty under Recall surfaces in tree-based and subspace-partition indices 
% even on easy data.

On mid-to-high LID datasets (AGNews, SimpleWiki, Gist, ImageNet), the gap between Recall and 1/Ratio QPS widens. Annoy fails to reach Recall@100~$\geq 0.95$ on any of the four datasets, but reaches 1/Ratio@100~$\geq 0.95$ at 
thousands of QPS on 
AGNews, SimpleWiki, and  Gist (ImageNet is unreachable under both metrics). For AGNews, the remaining algorithms reach $2\times$--$3\times$  speedups.
%of $3.06\times$ (HNSW), $2.27\times$ (RaBitQ), $2.44\times$ (SymphonyQG), and $1.99\times$ (SuCo). 
SimpleWiki is a notable exception for SymphonyQG. At such a high dimensionality ($D$\,=\,3{,}072), the product-quantization codes introduce high distortion such that graph traversal follows suboptimal paths, returning near-zero Recall across all configurations and preventing a speedup comparison. 
On Gist, HNSW and SuCo also fail to reach Recall@100~$\geq 0.95$ at any tested configuration, while 1/Ratio@100~$\geq 0.95$ is achievable for both at 5{,}338 and 1{,}821~QPS, respectively; RaBitQ and SymphonyQG reach speedups of $4.49\times$ and $5.27\times$. ImageNet produces the largest speedups of 1/Ratio over Recall despite moderate LID, reaching $9.98\times$ for HNSW, $8.21\times$ for RaBitQ, and $8.54\times$ for SymphonyQG; SuCo cannot reach Recall@100~$\geq 0.95$ here, and Annoy fails to reach either threshold. This suggests that while LID captures the main trend, 
factors such as the data distribution
% the embedding structure of individual datasets 
also contribute to the gap.
The pattern intensifies at higher thresholds. At $T = 0.99$, the average QPS speedup across reachable datasets rises to $4.38\times$ for HNSW, $3.90\times$ for RaBitQ, and $3.71\times$ for SuCo. Recall@100~$\geq 0.99$ becomes unreachable for HNSW on both Gist and ImageNet, while 1/Ratio@100~$\geq 0.99$ is reached in both. Most strikingly, Annoy cannot reach Recall@100~$\geq 0.99$ on any of the six tested datasets, while 1/Ratio@100~$\geq 0.99$ remains achievable in multiple configurations. Under 1/Ratio, these configurations achieve near-perfect distance-based quality; under Recall, they are rejected  because of 
low overlap to the true $k$NN set.
%identifier mismatch.

\begin{figure*}[t]
  \centering
  \includegraphics[width=\textwidth]{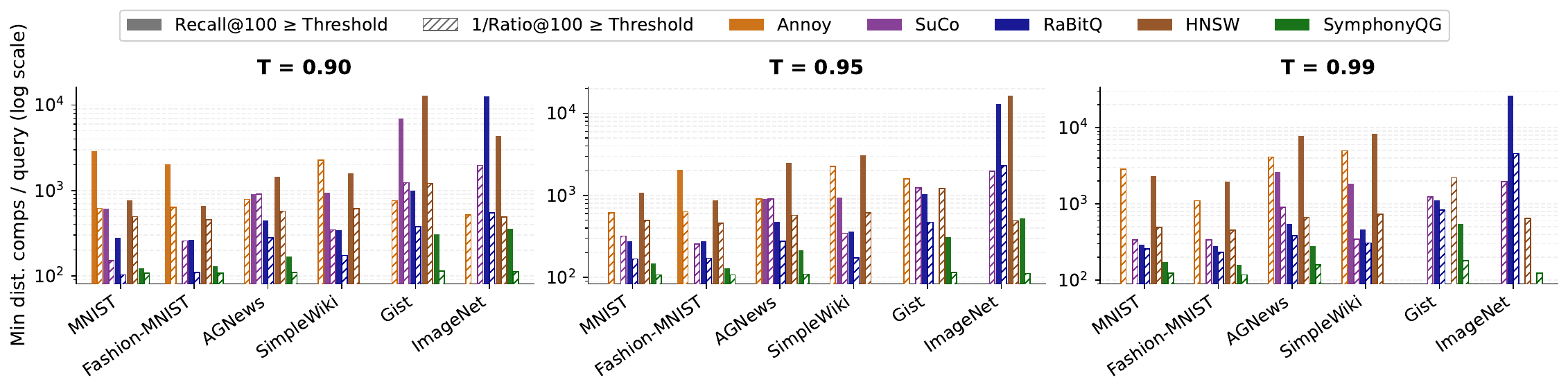}
  \caption{Minimum distance computations per query to reach quality threshold
  $T$ at $k = 100$, for $T \in \{0.90, 0.95, 0.99\}$ (rows) and each
  algorithm (columns). Missing bars: threshold unreachable.}
  \label{fig:distcomps}
  \vspace{-2mm}
\end{figure*}

\stitle{Distance computations (RQ1).}
QPS is system-dependent: it reflects CPU speed, memory bandwidth, SIMD width,
and parallelism. To isolate the algorithmic cost independently of hardware,
Figure~\ref{fig:distcomps} reports the minimum number of distance computations
per query needed to reach each quality threshold. For a given threshold $T$, we take the configuration with the fewest distance computations among all
configurations satisfying score~$\geq T$, and report this minimum separately
for Recall and 1/Ratio.
At $T = 0.95$ and $k = 100$, satisfying Recall requires more distance computations than satisfying 1/Ratio across all five algorithms: $9.36\times$ for HNSW, $2.48\times$ for RaBitQ, $2.38\times$ for SymphonyQG, $1.86\times$ for SuCo, and $3.22\times$ for Annoy
on average
% All numbers reported are medians 
across datasets. 
% On ImageNet at $T = 0.95$, HNSW requires $33.3\times$ more distance computations to satisfy Recall than 1/Ratio
% % , over an order of magnitude more algorithmic work 
% with no corresponding gain in distance quality. At $T = 0.99$, the HNSW ratio rises to $8.02\times$ and SuCo's to $4.14\times$, while Annoy cannot reach Recall@100~$\geq 0.99$ on any dataset.
The distance-computation gap and the QPS gap do not always align in magnitude,
because different algorithmic bottlenecks dominate in each case. For HNSW, the
distance-computation gap consistently exceeds the QPS gap (e.g., $33\times$
vs.\ $10\times$ on ImageNet at $T = 0.95$): graph-traversal overhead (pointer
chasing, branch misprediction, candidate-list management) adds a per-hop cost
that does not scale with the number of distances computed. For RaBitQ, the
pattern inverts: on Gist at $T = 0.99$, the distance-computation gap is
$1.33\times$ while the QPS gap is $4.57\times$, because each quantized
distance estimate is a cheap bitwise operation, so the distance-computation
count alone understates the total cost of scanning additional IVF clusters.
Distance computations and QPS therefore tell complementary stories, but both
independently confirm that 1/Ratio reaches quality thresholds with less work
than Recall. Note that this divergence is not an artifact of parallelism; the queries are independent, and running them across multiple cores speeds up both configurations by about the same factor, so the ratio between their throughputs is not affected.

\stitle{Index build time (RQ1).}
Next, we investigate whether the overhead of optimizing for Recall also extends to index construction. Table~\ref{tab:build_time} reports the ratio of the minimum build time required to satisfy Recall versus 1/Ratio at $k=100$. 
For some thresholds $T$, Recall $\ge T$ was unachievable; these cases were marked by N/A.

For HNSW, the build-time overhead depends on the threshold. At $T = 0.90$ and
$T = 0.95$, the cheapest Recall-satisfying and 1/Ratio-satisfying
configurations require essentially the same build time in most cases. On Gist the difference is exceptionally high at $T = 0.90$ ($18.39\times$); having high LID, Gist requires a much denser graph to satisfy Recall. At $T = 0.99$, however, the Recall-satisfying build cost rises on average to $2.73\times$,
% the 1/Ratio-satisfying cost: satisfying Recall@100~$\geq 0.99$ requires 
as denser graphs are required (higher $M$ or $\mathit{efConstruction}$) than 1/Ratio demands.
For SymphonyQG, the pattern is different: the average build-time ratio is $\approx2.40\times$ at $T = 0.90$ and $T = 0.95$, narrowing to $\approx1.06\times$ at $T = 0.99$ on three of the four reachable datasets. At the highest threshold, only the densest configurations satisfy either metric, so the build-time gap closes; Gist is an exception at $5.96\times$, where its high LID still demands a substantially denser graph for Recall@99 than for 1/Ratio@99.
For RaBitQ, SuCo, and Annoy, the minimum build time achieving each threshold is essentially the same under both metrics (average ratios of $1.00\times$, $1.37\times$ or less, and $1.00\times$ respectively), indicating that the cheapest Recall-satisfying and 1/Ratio-satisfying configurations use the same index construction parameters. For Annoy in particular, this comparison is restricted to $T \in \{0.90, 0.95\}$, as Recall@100~$\geq 0.99$ is unreachable on every tested dataset.

\begin{table}
\caption{Ratio of minimum index build time required to satisfy Recall@100~$\geq T$ versus 1/Ratio@100~$\geq T$. "N/A" indicates Recall~$\geq T$ was unreachable at any configuration. RaBitQ, SuCo, and Annoy are omitted as their build-time ratios stay at or near $1.00\times$ across all datasets and thresholds.}
\label{tab:build_time}
\centering
\resizebox{\columnwidth}{!}{%
\begin{tabular}{l rrr@{\hspace{4ex}} rrr}
\toprule
 & \multicolumn{3}{c@{\hspace{4ex}}}{\textbf{HNSW}} 
 & \multicolumn{3}{c}{\textbf{SymphonyQG}} \\
\cmidrule(r{4ex}){2-4} \cmidrule{5-7}
Dataset & $T=0.90$ & $T=0.95$ & $T=0.99$ & $T=0.90$ & $T=0.95$ & $T=0.99$ \\
\midrule
MNIST           & $1.00\times$ & $1.00\times$ & $1.44\times$ & $2.28\times$ & $2.28\times$ & $1.09\times$ \\
Fashion-MNIST   & $1.00\times$ & $1.00\times$ & $1.00\times$ & $2.13\times$ & $2.12\times$ & $1.08\times$ \\
ImageNet        & $1.84\times$ & $3.63\times$ & N/A & $2.31\times$ & $2.31\times$ & N/A \\
AGNews         & $1.00\times$ & $1.00\times$ & $4.58\times$ & $2.82\times$ & $2.82\times$ & $1.00\times$ \\
SimpleWiki      & $1.00\times$ & $1.52\times$ & $3.91\times$ & N/A & N/A & N/A \\
Gist            & $18.39\times$ & N/A & N/A & $2.45\times$ & $2.45\times$ & $5.96\times$ \\
\bottomrule
\end{tabular}%
}
\end{table}

\begin{comment}
\begin{table}[t]
\centering
\caption{Search phase memory (MB) at target quality thresholds $T \in \{0.90, 0.95, 0.99\}$. Single values denote identical footprints for both metrics; divergences are formatted as \textit{Recall} / \textit{1/Ratio}.}
\label{tab:memory}
\begin{tabular}{lrrr}
\toprule
\textbf{Dataset} & \textbf{HNSW} & \textbf{RaBitQ} & \textbf{SymphonyQG} \\
\midrule
\multicolumn{4}{c}{\textbf{Thresholds $T \in \{0.90, 0.95\}$}} \\
\midrule
MNIST               & 453           & 457   & 952 / 951 \\
Fashion-MNIST       & 443           & 466   & 868 \\
AG News             & 6,429         & 6,538 & 11,975 / 11,974 \\
SimpleWiki          & 6,460         & 6,673 & N/A \\
Gist ($T=0.90$)     & 7,967 / 7,839 & 6,903 & 15,042 \\
Gist ($T=0.95$)     & N/A / 7,839   & 6,903 & 15,042 \\
ImageNet ($T=0.90$) & 6,758         & 6,811 & 15,985 \\
ImageNet ($T=0.95$) & 7,086 / 6,758 & 6,811 & 15,985 \\
\midrule
\multicolumn{4}{c}{\textbf{Threshold $T=0.99$}} \\
\midrule
MNIST         & 453           & 457   & 953 / 952 \\
Fashion-MNIST & 443           & 466   & 868 \\
AG News       & 6,528 / 6,429 & 6,538 & 11,975 \\
SimpleWiki    & 6,493 / 6,460 & 6,673 & N/A \\
Gist          & N/A / 7,839   & 6,903 & 19,908 / 15,042 \\
ImageNet      & N/A / 6,758   & 6,811 & N/A / 15,985 \\
\bottomrule
\end{tabular}
\end{table}
\end{comment}

\stitle{Memory footprint (RQ1).}
%Across most algorithms, datasets, and thresholds, the average ratio of Recall-satisfying to 1/Ratio-satisfying memory is near $1.00\times$ (exactly $1.00\times$ for RaBitQ and Annoy). This flatness reflects that query-time memory is dictated by base index parameters, not search-time routing budgets. Memory overheads emerge only when high Recall targets force denser index construction. The most striking example is SymphonyQG on Gist at $T=0.99$, where Recall demands $1.32\times$ the memory of 1/Ratio, followed by HNSW on ImageNet at $T=0.95$ ($1.05\times$ overhead). Importantly for graph-based indices, construction-time spikes do not translate to equivalent memory inflation. On Gist at $T=0.90$, high LID forces HNSW to require $18.39\times$ more build time to satisfy Recall, yet the final memory footprint increases by only $1.6\%$. While the Recall-satisfying configuration doubles the graph connectivity parameter from $M=16$ to $M=32$, search-phase memory is dominated by the raw vector data~\cite{DBLP:journals/pami/MalkovY20, DBLP:conf/icde/YueX0TL24}.
We also studied the memory overhead of indices optimized for Recall vs. 1/Ratio. We report the main observations; details are not shown due to the interest of space.
Across most algorithms, datasets, and thresholds, the average ratio of Recall-satisfying to 1/Ratio-satisfying memory is near $1.0\times$ (exactly $1.00\times$ for RaBitQ and Annoy, and within a few percent for SuCo, HNSW, and SymphonyQG). Memory overheads become visible only at the highest quality thresholds, and even there remain modest: the largest gap we observe is SymphonyQG on Gist at $T=0.99$, where Recall demands $1.32\times$ the memory of 1/Ratio, followed by HNSW on ImageNet at $T=0.95$ ($1.05\times$ overhead). Interestingly, large build-time gaps do not translate into large memory gaps. The most striking case is HNSW on Gist at $T=0.90$: the Recall-satisfying configuration doubles the graph connectivity parameter from $M=16$ to $M=32$ and requires $18.39\times$ more build time, yet the final memory footprint grows by only $1.6\%$. This is consistent with the in-memory footprint of HNSW being dominated by raw vector storage~\cite{DBLP:journals/pami/MalkovY20, DBLP:conf/icde/YueX0TL24} rather than graph adjacency.

\stitle{Algorithm rankings (RQ2).}
% Replacing Recall with 1/Ratio increases the maximum achievable QPS for all five algorithms, but by an algorithm-dependent amount rather than a uniform one. 
As shown in Figure~\ref{fig:qps_bars}, the relative ranking of algorithms in terms of throughput remains largely unchanged if we optimize them for 1/Ratio instead of Recall: SymphonyQG generally achieves the highest QPS under both metrics across most datasets, with HNSW, RaBitQ, SuCo, and Annoy trading the remaining positions depending on the dataset and threshold. 
The relative gaps between algorithms, however, are not preserved;
%: because the QPS gain from switching metrics varies across algorithms (e.g., on MNIST at $T=0.95$, $1.00\times$ for HNSW versus $2.19\times$ for SymphonyQG), 
cross-algorithm differences can widen or narrow. At $T=0.95$, SymphonyQG is $3.1\times$ faster than HNSW on MNIST under Recall but $6.9\times$ faster under $1/\text{Ratio}$; on Fashion-MNIST, HNSW is $2.9\times$ faster than Annoy under Recall but only $1.4\times$ faster under $1/\text{Ratio}$.
%The small re-orderings that do occur involve 
Algorithms with near-identical throughput (e.g., HNSW and RaBitQ on SimpleWiki at $T=0.95$) may change rank if 1/Ratio is used instead of Recall, but this does not
change the winner algorithm.
%out of first place. 

The same observation holds along the other cost axes. For minimum distance computations per query (Figure~\ref{fig:distcomps}), satisfying Recall requires between $1.86\times$ and $9.36\times$ more distance computations than 1/Ratio across the five algorithms at $T=0.95$, a multiplier consistent enough not to reorder them.
% : the quantisation-based method RaBitQ and the graph-quantisation hybrid SymphonyQG remain the cheapest in absolute terms, while HNSW and Annoy remain the most expensive method, under either metric. 
Regarding index build time, the Recall-to-1/Ratio ratio is essentially flat for RaBitQ, SuCo, and Annoy and reaches up to $2.39\times$ on average for SymphonyQG at $T=0.95$, but no algorithm changes its rank under the new metric. Finally, search-phase memory is essentially identical under both metrics for every algorithm, so cross-algorithm memory rankings are unchanged. 

The overall conclusion is therefore: changing the evaluation metric from Recall to 1/Ratio shifts the absolute cost at which each algorithm reaches operational quality, but it does not change which algorithms are faster, lighter, or cheaper to build. This has a practical implication: since the ranking is stable, the comparative conclusions of Recall-based benchmarks \cite{DBLP:journals/is/AumullerBF20, DBLP:journals/corr/abs-2505-17810, DBLP:journals/pacmmod/LiYLZCM25} carry over to $1/\text{Ratio}$, and can be reproduced more cheaply by benchmarking under $1/\text{Ratio}$. Only the magnitude of the differences is metric-specific.

\begin{figure*}[t]
  \centering
  \includegraphics[width=\textwidth]{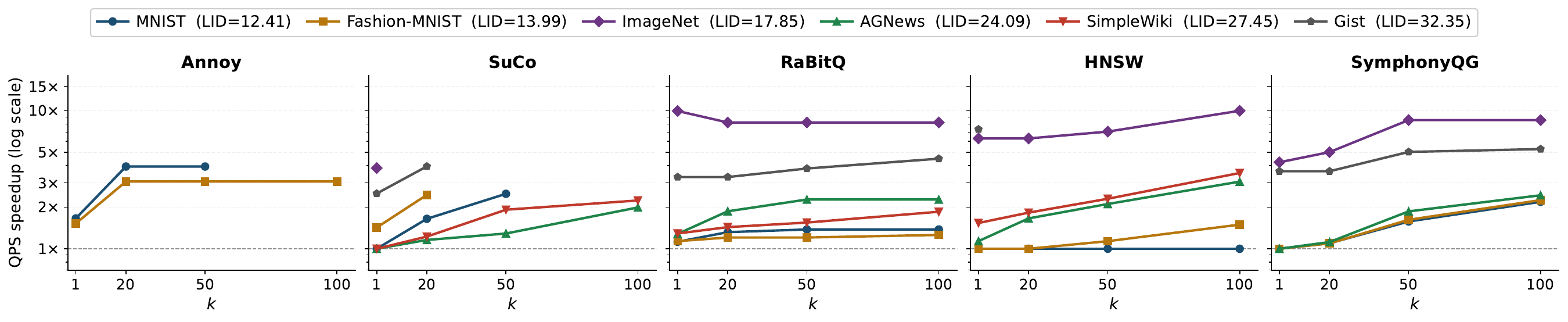}
  \caption{QPS speedup of 1/Ratio over Recall as a function of $k \in \{1, 20, 50, 100\}$ at $T=0.95$, per algorithm and dataset. Absent lines indicate that Recall@$k \geq 0.95$ was unreachable at all tested configurations: at $T=0.95$, Annoy maintains Recall across the full range of $k$ only on Fashion-MNIST, SuCo only on AGNews and SimpleWiki, and HNSW on Gist is reachable only at $k=1$.}
  \label{fig:k_analysis}
\end{figure*}

\stitle{Effect of $k$ (RQ3).}
Figure~\ref{fig:k_analysis} reports the QPS speedup of 1/Ratio over Recall at $T = 0.95$ as a function of $k \in \{1, 20, 50, 100\}$, broken down by algorithm and dataset. The speedup grows monotonically with $k$ in most algorithm-dataset-threshold combinations we tested. As $k$ increases, the query boundary in the search space that encloses the true NNs expands, and the probability that an ANN algorithm returns a neighbor geometrically as close as a ground-truth vector but with a different identifier rises~\cite{DBLP:conf/icdt/BeyerGRS99}, penalizing Recall while leaving 1/Ratio essentially unchanged. On ImageNet under HNSW, the QPS speedup rises from $6.29\times$ at $k = 1$ to $9.98\times$ at $k = 100$; on Gist under SymphonyQG, from $3.64\times$ at $k = 1$ to $5.27\times$ at $k = 100$. The same trend holds at $T = 0.90$ and $T = 0.99$, with larger absolute speedups at higher thresholds.

\begin{figure*}[t]
\centering
\includegraphics[width=\linewidth]{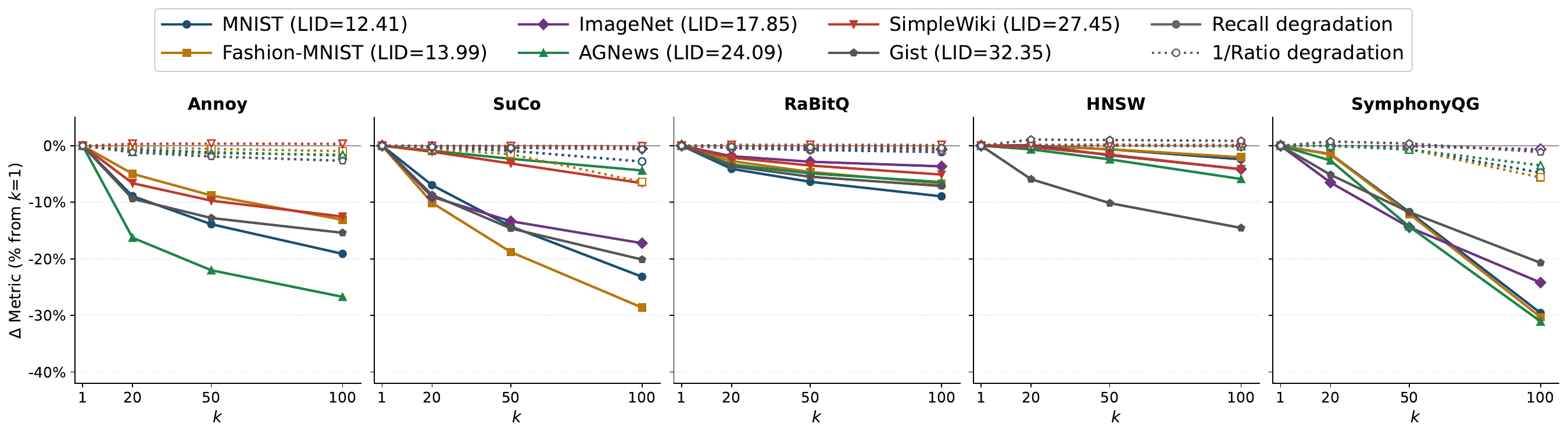}
\caption{Average change in Recall@$k$ and 1/Ratio@$k$ relative to their $k{=}1$ values, as $k$ grows from 1 to 100, across all parameter configurations per algorithm. Recall degrades substantially with $k$, while 1/Ratio remains nearly flat.}
\label{fig:lid_degradation}
\vspace{-2mm}
\end{figure*}

\stitle{Geometric divergence across datasets (RQ3).}
The results above show that the efficiency gap between Recall and 1/Ratio varies across datasets and is related to, though not fully determined by, LID. To isolate this relationship, we examine how each metric degrades as $k$ grows, fixing the algorithm configuration and varying only $k \in \{1, 20, 50, 100\}$. Figure~\ref{fig:lid_degradation} reports, for every algorithm and dataset, the average percentage change of Recall@$k$ and 1/Ratio@$k$ relative to their $k{=}1$ values across all configurations in the parameter sweep. %Datasets are colored by intrinsic dimensionality (LID), from MNIST to Gist.
A consistent pattern emerges across all five algorithms: Recall@$k$ drops substantially as $k$ grows, with the magnitude broadly tracking LID for RaBitQ, HNSW, and SymphonyQG. For Annoy and SuCo, the Recall@$k$ degradation is less related to LID; because tree-based and subspace-partitioning indices restrict query routing to pre-computed structures (such as leaf nodes or projected subspaces), their approximation errors are dominated by structural partitioning boundaries rather than pure geometric distance spacing \cite{DBLP:conf/cvpr/BabenkoL12, DBLP:conf/bigdataconf/YanWWWL18}.
On the other hand,
% (for HNSW, Gist loses about 15\% by $k{=}100$ while MNIST and Fashion-MNIST stay within 1\%), whereas 
$1/\text{Ratio}@k$ stays within a much narrower band, with $y$-axis ranges roughly 4$\times$-5$\times$ smaller. 
% For HNSW, $1/\text{Ratio}$ stays within $\pm1\%$ of its $k{=}1$ value on every dataset; for RaBitQ, SuCo, and Annoy the maximum drop is 2-3\%. 
SuCo exhibits the largest $1/\text{Ratio}$ drop (up to $6.4\%$), which is still much smaller than its corresponding Recall drop ($28.6\%$).

The connection to dimensionality is geometric. 
In high-LID spaces the data manifold is locally uniform, so successive neighbors are nearly equidistant; as $k$ grows, the retrieved set extends deeper into this equidistant shell, making strict identifier matching (what Recall measures) increasingly unreliable while the distance quality of retrieved vectors (what $1/\text{Ratio}$ measures) stays high. 
In low-LID spaces, the effect of $k$ is smaller, %successive neighbours are well-separated, exact identifiers are more distinguishable, and the two metrics agree. This 
explaining the broader trend in Figure~\ref{fig:lid_degradation}.
%, though the relationship is not strictly monotone, as dataset-specific structure can amplify or dampen the effect independently of LID.

%%%%%%%%%%%%%%%%%%%%%%%%%%%%%%%%%%%%%%%%%%%%%%%%%%%%%%%%%%%%%%%%%%%%%%%%%%%%%%%%%%%%%%%%%%%%%%%%

\begin{figure*}[t]
  \centering
  \includegraphics[width=\linewidth]{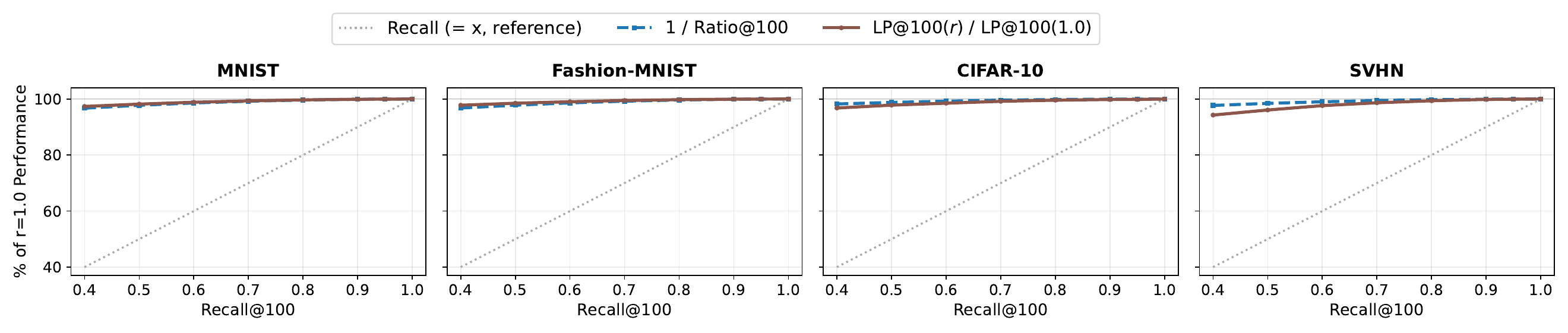}
    \caption{LP@100 and 1/Ratio@100 (normalized to exact search) versus
        synthetic recall.  Both metrics stay near 1.0 as recall drops
        to 0.4; the dotted $y{=}x$ diagonal shows how far Recall
        diverges from actual downstream quality.}
  \label{fig:lp_stability}
\end{figure*}

\subsection{Classification (RQ4)}
\label{sec:classification}

The cost-accuracy experiments show that 1/Ratio@$k$ reaches operational thresholds at substantially lower cost than Recall@$k$.  But a metric with easy-to-achieve quality targets,  like 1/Ratio@$k$, is only useful if it actually tracks what matters: downstream task quality.  In this subsection we test this directly
by investigating whether 1/Ratio@$k$ is aligned with
downstream task retrieval quality.
% .  If a quality metric is a faithful proxy for task performance, its variation range should match that of the task metric as retrieval degrades, a proxy that swings far more (or far less) than the task metric is miscalibrated. 
We use $k$-nearest neighbor classification on four image benchmarks (MNIST, Fashion-MNIST, CIFAR-10, and SVHN) and measure Label Precision at $k = 100$ (LP@100), the fraction of retrieved vectors that share the query's class label.  We ask two questions: (i)~does LP@100 actually degrade as Recall drops, and (ii)~which retrieval metric, Recall or 1/Ratio tracks LP@100 more faithfully?
%To systematically evaluate how downstream classification accuracy responds to recall loss, we rely on the synthetic substitution protocol defined in Section~\ref{sec:setup}. However, to meaningfully compare these results across diverse datasets and against the Recall metric itself, metric normalisation is required. %Label Precision (LP) reflects a dataset's absolute difficulty rather than its relative retrieval degradation; for instance, a perfect exact search yields an LP@100 of approximately 0.887 on MNIST but only 0.228 on the more complex CIFAR-10 dataset. 
%Without normalisation, LP, 1/Ratio and Recall inhabit completely different scales, making cross-dataset comparisons and direct metric evaluations effectively impossible. 
%By defining normalized precision as $LP_{norm} = \text{LP}@100(r) / \text{LP}_{\text{exact}}$, we establish a universal axis where 1.0 indicates perfect exact-search performance. This alignment eliminates dataset-specific variations, providing a unified scale that allows preserved downstream quality to be directly compared against the Recall metric.
%To systematically evaluate how downstream classification accuracy responds to recall loss, 
To answer, we apply the synthetic recall control and normalization 
%defined in Section~\ref{sec:setup} 
and report normalized precision $LP_{norm}$, as defined in Section~\ref{sec:setup}.

\stitle{Downstream quality under recall degradation.}
Figure~\ref{fig:lp_stability} plots LP@100 and 1/Ratio@100
against synthetic recall level for all four datasets, both normalized
so that the quality of exact search ($r = 1.0$)  corresponds to 1.0. As recall drops from 1.0 to 0.4 (a 60\% change) we observe a consistent behavior across all datasets: downstream quality barely moves.
At $r = 0.4$, $LP_{norm}$ remains remarkably stable at 0.974 for MNIST, 0.978 for Fashion-MNIST, 0.968 for CIFAR-10, and 0.943 for SVHN. Over the exact same sweep, 1/Ratio@100 perfectly mirrors this stability, resting at 0.967, 0.968, 0.982, and 0.977 respectively. Even in the worst-case dataset (SVHN), $LP_{norm}$ only drops by 0.057 (from 1.0 down to 0.943). This 
%absolute drop 
is an order of magnitude smaller than the massive 0.60 drop in Recall that produced it.
The dotted $y = x$ diagonal in each plot of Figure~\ref{fig:lp_stability} marks where Recall itself sits as a quality predictor. The gap between the diagonal and the LP and 1/Ratio curves is wide and consistent: LP and 1/Ratio both stay above 0.94 across the entire sweep, while Recall drops to a level (0.4)  %considered unacceptable for
%, i.e., a level that
% . The two quality curves move together; Recall moves alone, in a direction that has no correspondence 
% that does not correspond to the 
far from the actual  task quality.

\begin{figure}[htb]
    \centering

    \includegraphics[width=0.45\textwidth]{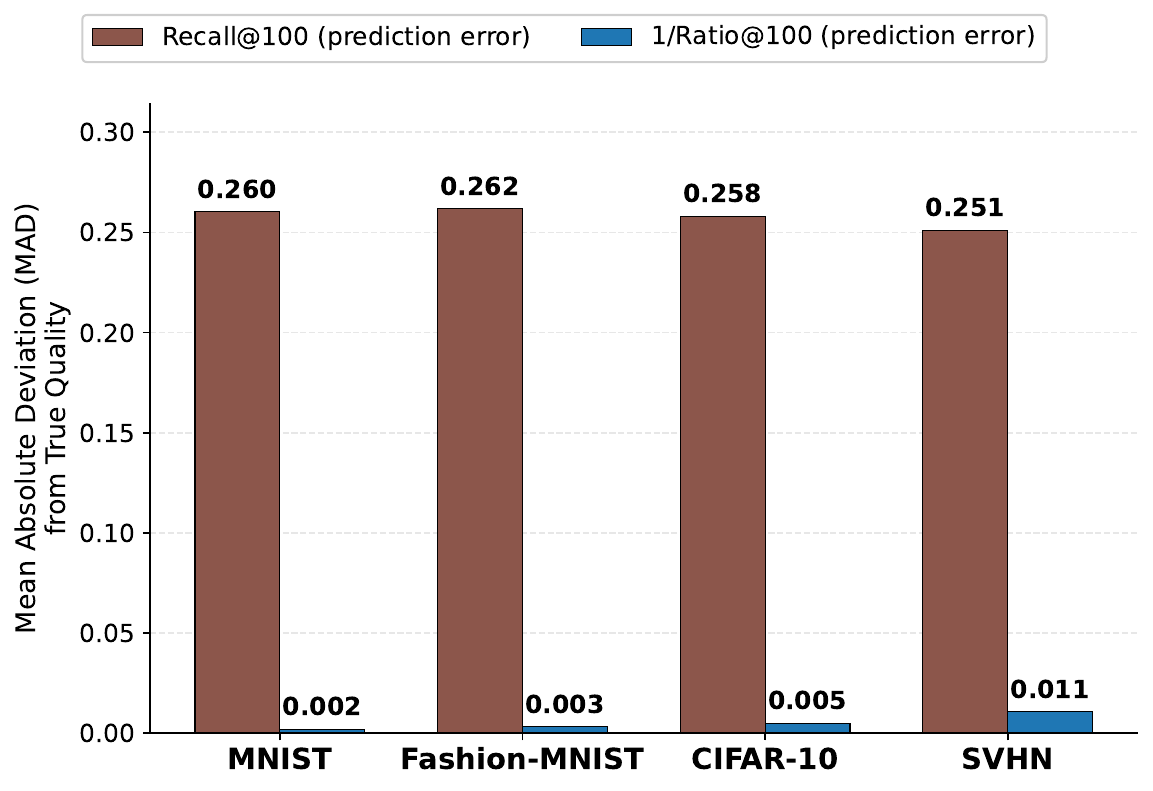}
    \caption{
        Mean Absolute Deviation (MAD) of Recall and $1/\text{Ratio}$ from the true normalized downstream quality (LP@100). Perfect proxies for quality have zero MAD.
    }
    \label{fig:metric_correlation}
    % \vspace{-4mm} % Adjust spacing to fit tightly with the rest of the text
\end{figure}

\stitle{Quantifying the gap between the two metrics.}
To measure how well each metric predicts downstream performance, we compute the Mean Absolute Deviation (MAD) between the predictor metric (Recall or 1/Ratio) and the true normalized quality ($LP_{norm}$) across the $r \in [0.4, 1.0]$ sweep. If a metric were a perfect proxy for quality, its value would exactly equal the normalized LP@100, yielding a MAD of zero.
As Figure~\ref{fig:metric_correlation} demonstrates, Recall is a remarkably poor predictor of actual quality, heavily underestimating performance with MAD values around 0.26.
%. Its MAD values are 0.260 (MNIST), 0.262 (Fashion-MNIST), 0.258 (CIFAR-10), and 0.251 (SVHN). For example, a Recall of 0.40 implies that only 40\% of quality is preserved, but in reality, $LP_{norm}$ is near 0.94 across all datasets. Recall underestimates performance by over 25\% on average. 
In contrast, $1/\text{Ratio}$'s MAD is up to two orders of magnitude lower.
%0.002 (MNIST), 0.003 (Fashion-MNIST), 0.005 (CIFAR-10), and 0.011 (SVHN). 
For instance, on MNIST at $r=0.4$, a 1/Ratio value of 0.967 closely matches the actual $LP_{norm}$ of 0.974, a prediction error of less than 1\%. 
Overall, besides enabling more efficient ANN search configurations, 
1/Ratio serves as a much better predictor of downstream quality compared to Recall.
%does not.
%\nikos{this experiment is redundant, as low MAD values are evident from Figure 5, already.}

\stitle{Stability of the predictor metric across neighborhood sizes.}
The experimental results presented thus far establish that at $k=100$  $1/\text{Ratio}$ serves as a far more accurate proxy for downstream quality ($LP_{norm}$) than Recall. To ensure this advantage is not an artifact of evaluating a wide search depth (large $k$), Table~\ref{tab:multi_k_range} reports the MAD of each metric against $LP_{norm}$ for smaller neighborhood sizes ($k \in \{1, 20, 50, 100\}$).
The disconnect between Recall and downstream task performance holds at every search depth. Recall's MAD remains high, ranging between 23.3\% and 26.6\% across all datasets and $k$ values. Regardless of neighborhood size, Recall consistently overstates the drop in true classification quality.
By contrast, 1/Ratio is a more accurate predictor even at $k=1$. In this case, where substituting a single vector determines the prediction, its MAD is between 1.0\% and 2.3\%. As $k$ increases, this advantage widens. 
%At $k=100$, the MAD for $1/\text{Ratio}$ decreases further to 0.2\%-1.1\%.
% The effectiveness of $1/\text{Ratio}$ as a quality predictor is not limited to large neighbourhood sizes. Recall's inability to track downstream quality persists at every $k$, whereas $1/\text{Ratio}$ remains accurate and improves as the search depth expands.

\begin{table}[htb]
\centering
\caption{%
  Mean Absolute Deviation (MAD), representing the prediction error (in \%) of Recall and $1/\text{Ratio}$ from normalized downstream quality ($LP_{norm} = \text{LP}@k\,/\,\text{LP}@k(1.0)$), for $k \in \{1,20,50,100\}$.
  Lower values indicate a more faithful proxy for true downstream quality.
}
\label{tab:multi_k_range}
\resizebox{\columnwidth}{!}{%
\begin{tabular}{llcccc}
\toprule
 & & \multicolumn{4}{c}{\textbf{Prediction error (\%)}} \\
\cmidrule(lr){3-6}
\textbf{Dataset} & \textbf{Predictor} &
  $\mathbf{k=1}$ & $\mathbf{k=20}$ & $\mathbf{k=50}$ & $\mathbf{k=100}$ \\
\midrule
\multirow{2}{*}{MNIST}
  & Recall              & 26.6 & 26.5 & 26.3 & 26.0 \\
  & $1/\text{Ratio}$    &  1.3 & 0.6 &  0.4 &  0.2 \\
\midrule
\multirow{2}{*}{Fashion-MNIST}
  & Recall              & 26.0 & 26.4 & 26.4 & 26.2 \\
  & $1/\text{Ratio}$    &  1.0 & 0.4 &  0.4 &  0.3 \\
\midrule
\multirow{2}{*}{CIFAR-10}
  & Recall              & 23.3 & 25.6 & 25.8 & 25.8 \\
  & $1/\text{Ratio}$    &  2.3 & 0.8 &  0.6 &  0.5 \\
\midrule
\multirow{2}{*}{SVHN}
  & Recall              & 24.5 & 25.3 & 25.2 & 25.1 \\
  & $1/\text{Ratio}$    &  1.3 & 1.0 &  1.0 &  1.1 \\
\bottomrule
\end{tabular}%
}
\end{table}

\subsection{RAG Experiments (RQ4)}
\label{sec:rag}

The classification experiments show that Label Precision remains stable as Recall degrades, and that $1/\text{Ratio}@k$ tracks this stability far more faithfully than Recall does. 
We now extend the analysis to retrieval-augmented generation (RAG), where the downstream 
task is fundamentally more complex: the system must synthesize a free-form answer 
from retrieved passages and quality is measured by how closely that answer matches 
a reference. 

% \nikos{before talking about the datasets, experimental methodology should be described more concretely}

%We evaluate across five datasets under two reference paradigms. First, on three QA benchmarks with human-written gold answers (HotpotQA, MS-MARCO, and PubMedQA), we measure absolute task accuracy. For each query and simulated recall level, the LLM generates an answer using the approximate retrieved context, and we evaluate this generated text directly against the independent, human-written ground-truth references.
%Next, we evaluate the LLM's internal consistency across two BEIR benchmarks (SciFact and NFCorpus). Again, for each query and simulated recall level, the LLM generates an answer from the approximate context, compared against a temperature-0.3 reference from the exact top-$k$. This prevents the deterministic self-comparison artifact seen at temperature 0, where the model reproduces the $r=1.0$ answer without any change. 

We evaluate across five datasets under two reference paradigms, as discussed in Section \ref{sec:setup}.
First, on three QA benchmarks with human-written gold answers (HotpotQA, MS-MARCO, and PubMedQA), we measure absolute task accuracy. For each query and synthetic recall level, the LLM generates an answer from the approximate retrieved context, and we evaluate this generated text against the independent, human-written ground-truth reference.
Next, we evaluate the LLM's internal consistency on two BEIR benchmarks (SciFact and NFCorpus), where no human gold answers are available. Again, for each query and synthetic recall level, the LLM generates an answer from the approximate context at temperature 0. The difference is the reference, which is now generated by the same LLM from the exact top-$k$ neighbors at temperature 0.3, to avoid the deterministic self-comparison artifact that temperature 0 would cause at $r=1.0$. %likely to be a lil confusing like this. i explain the whole temperature thing later on. 
These experiment setups isolate the generation phase, allowing us to accurately quantify how approximation impacts the final output; specifically, whether supplying the LLM with geometrically close near-misses instead of exact nearest neighbors degrades task accuracy or alters the synthesized response.

\begin{figure*}[htbp]
    \centering
    \includegraphics[width=0.99\textwidth]{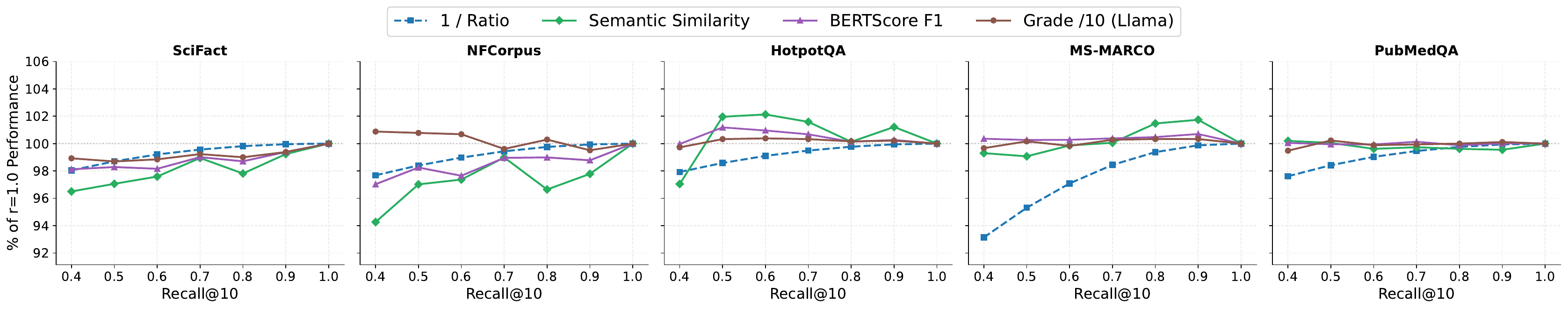}
    \vspace{-1mm}
    \caption{Downstream generation quality (Semantic Similarity, BERTScore F1,
    Grade/10) and geometric stability ($1/\text{Ratio}$) versus synthetic
    search recall, normalized to the $r{=}1.0$ exact-search baseline.}
    \label{fig:all_combined}
\end{figure*}

\stitle{Answer quality under recall degradation.} We begin with the human-annotated benchmarks, where the gold answer is entirely independent of the retrieval and generation pipeline. Figure~\ref{fig:all_combined} plots the three normalized quality metrics and $1/\text{Ratio}@10$ against synthetic recall for the three datasets. Answer quality is effectively decoupled from recall across the full range $r \in [0.4, 1.0]$: BERTScore F1 varies by at most 1.2\% on HotpotQA and under 1\% on MS-MARCO and PubMedQA. Similarly, the LLM-graded score varies by under 1\% on all three datasets. 
Semantic Similarity is the noisiest of the three metrics (up to 5.1\% variation across recall levels). On HotpotQA, it is non-monotonic: it starts at 97.0\% at recall $r = 0.4$, but peaks at 102.1\% at $r = 0.6$, surpassing the perfect-recall baseline. 

The spikes above $100\%$ do not mean the answer actually got better; rather, the mix of passages at that specific step triggers slightly different wording that happens to align a bit closer with the human text. While these wording shifts make language scores look messy, the underlying system is actually highly stable. We can verify this through the $1/\text{Ratio}$ metric, which bypasses text entirely to measure the vector distances. Across all three datasets, $1/\text{Ratio}$ remains tight, dropping by at most 6.9\% on MS-MARCO and under 2.5\% on HotpotQA and PubMedQA. Because the text quality drops even less than the vector math, $1/\text{Ratio}$ serves as a safe, conservative floor for how well the system performs. 

\begin{comment}
\begin{figure*}[htbp]
    \centering
    \includegraphics[width=0.95\textwidth]{part2_combined.pdf}
    \caption{Downstream generation quality metrics and geometric stability ($1/\text{Ratio}$) versus synthetic search recall for HotpotQA, MS-MARCO, and PubMedQA.}
    \label{fig:part2_combined}
\end{figure*}
\end{comment}

Table~\ref{tab:raw_metrics_combined} reports the exact raw scores ranges across all configurations, confirming this flat performance trend. On HotpotQA, MS-MARCO, and PubMedQA, the text quality metrics barely budge. For example, on PubMedQA, the LLM-graded score varies by a mere $0.07$ points on a 10-point scale across the entire recall sweep. Recall forces a massive drop of $60\%$ (from $1.0$ down to $0.4$). In contrast, $1/\text{Ratio}$ numbers remain remarkably steady and 
never drop below $93.14\%$ on any dataset. Ultimately, these raw values confirm that configurations that Recall flags as ``inadequate'' can still produce answers of identical quality as exact $k$NN search.

\begin{table}[htbp]
\centering
\caption{Range of raw quality metrics across the recall sweep $r \in [0.4, 1.0]$ for all five RAG datasets.}
\label{tab:raw_metrics_combined}
\setlength{\tabcolsep}{4pt} % default is 6pt
\begin{tabular}{lcccc}
\toprule
\textbf{Dataset} & \textbf{$1/\text{Ratio}$} & \textbf{Sem. Sim.} & \textbf{BERTScore} & \textbf{Grade/10} \\
\midrule
HotpotQA & 0.98--1.00 & 0.52--0.55 & 0.75--0.76 & 9.13--9.19 \\
MS-MARCO & 0.93--1.00 & 0.51--0.53 & 0.76 & 8.95--9.01 \\
PubMedQA & 0.98--1.00 & 0.75--0.76 & 0.80 & 8.66--8.73 \\
SciFact  & 0.98--1.00 & 0.80--0.83 & 0.86--0.87 & 8.80--8.92 \\
NFCorpus & 0.98--1.00 & 0.76--0.81 & 0.81--0.84 & 8.71--8.83 \\
\bottomrule
\end{tabular}
\end{table}

\begin{comment}
\begin{figure}[htbp]
    \centering
    \includegraphics[width=0.45\textwidth]{fig_part1_gap_normalized.pdf}
    \caption{Part 1: Quality vs. Recall@10 (normalized to R=1.0 baseline) for BEIR datasets.}
    \label{fig:part1_gap_normalized}
\end{figure}
\end{comment}

\stitle{LLM internal consistency under approximate retrieval.} The human-annotated experiments measure quality against an external ground truth. A complementary question is whether the LLM's generative process is itself sensitive
to recall degradation, i.e., whether substituting exact neighbors with geometrically close but identifier-distinct passages causes the model's output to meaningfully diverge from what it would produce exact $k$NN retrieval.

We test this on two BEIR datasets (SciFact and NFCorpus). For each query, a reference answer is generated from the exact $k$ nearest neighbors using Llama~3.1:8b at temperature~0.3, while the sweep answers ($r \in [0.4, 1.0]$) are generated at temperature~0 and evaluated against this reference. Temperature~0.3 avoids a deterministic artifact: at temperature~0 the $r{=}1.0$ sweep answer and the reference are identical strings, inflating apparent degradation at lower recall levels. Temperature~0.3 introduces slight wording variance, providing a high-quality but non-identical string. 

Figure~\ref{fig:all_combined} normalizes all metrics to their $r{=}1.0$ values, so that 100\% represents performance under perfect recall and any drop reflects recall-induced degradation alone. The degradation is remarkably small: Semantic Similarity drops 3.5\% on SciFact and 5.7\% on NFCorpus as recall degrades to $r{=}0.4$; BERTScore~F1 drops 2--3\%; the LLM-graded score drops 1.5\% or less. Over the same range, $1/\text{Ratio}@10$ drops by under 2.4\% on both datasets. BERTScore and the LLM grade closely track the geometric signal; Semantic Similarity shows slightly more variation (3.5\% on SciFact, 5.7\% on NFCorpus). In all cases, quality loss is bounded within a narrow interval even when 
%over a range where 
Recall degrades by 60\%.
%, confirming that identifier overlap is a poor indicator for
% approximate retrieval quality.

\begin{figure}[htbp]
    \centering
    \includegraphics[width=0.45\textwidth]{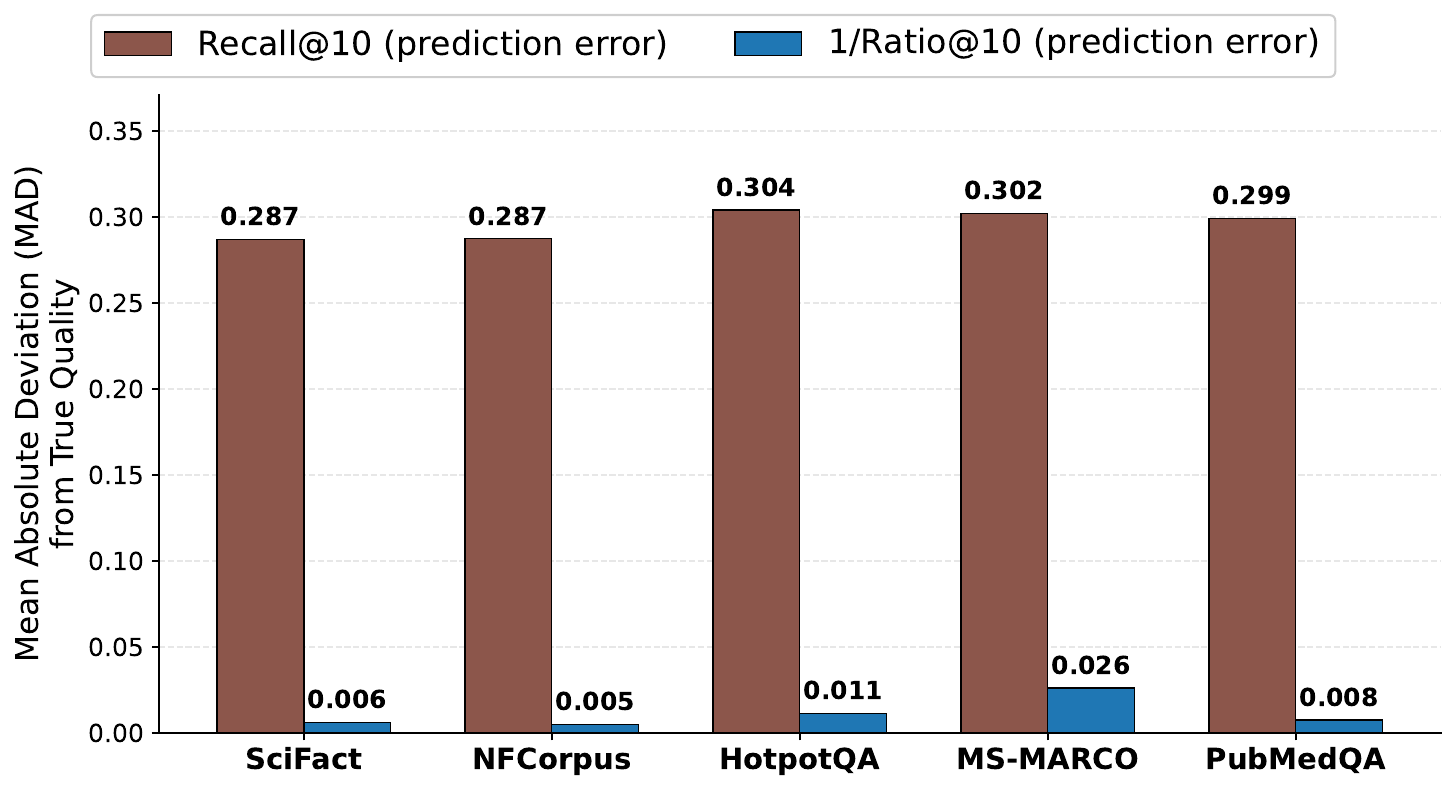}
    \caption{MAD of Recall and 1/Ratio from true RAG quality across datasets.}
    \label{fig:mad_comparison}
\end{figure}

\stitle{Recall vs.\ $1/\text{Ratio}$ as quality proxies.}
%Both experiments arrive at the same conclusion: answer quality is quite stable across a wide range of Recall levels. 
%The question is whether 1/Ratio reflects this stability. To quantify this, we compute its MAD against a composite true quality score (the average of the three normalized metrics).
% As Recall swings 60\% across $r \in [0.4, 1.0]$, 
% %a variation no downstream metric reflects. 
% $1/\text{Ratio}@10$ degrades by at most 6.9\% on MS-MARCO, under 2.5\% on HotpotQA and PubMedQA, and under 2.4\% on both BEIR datasets. On MS-MARCO, $1/\text{Ratio}$ drops 6.9\% while BERTScore~F1 changes by under 0.3\% and the LLM grade by under 1\%. BERTScore and the LLM grade degrade less than $1/\text{Ratio}$ in every tested dataset, confirming that $1/\text{Ratio}$ is a conservative proxy that overstates the true quality cost of approximation.
%Figure~\ref{fig:mad_comparison} shows the MAD of Recall and 1/Ratio across 
%makes this precise. On 
%the human-annotated benchmarks: $1/\text{Ratio}$'s MAD from the true normalized quality is 1.1\% (HotpotQA), 2.6\% (MS-MARCO), and 0.8\% (PubMedQA); Recall's MAD is approximately 30\% across
%all three -an improvement of $11\times$ to $37\times$. On the BEIR benchmarks, $1/\text{Ratio}$ achieves a MAD of 0.6\% (SciFact) and 0.5\% (NFCorpus), compared to Recall's 28.7\%---a $47\times$ to $57\times$ improvement. Across all five datasets, $1/\text{Ratio}$ is consistently a much better predictor of downstream RAG quality than Recall.
%\stitle{Recall vs.\ $1/\text{Ratio}$ as quality proxies.}
Both experiments arrive at the same conclusion: answer quality is quite stable across a wide range of Recall levels. To quantify whether 1/Ratio reflects this stability better than Recall, we compute their MAD against a composite true quality score (the average of the three normalized metrics). Figure~\ref{fig:mad_comparison} shows the MAD of Recall and $1/\text{Ratio}$ across the human-annotated benchmarks: $1/\text{Ratio}$'s MAD is 1.1\% (HotpotQA), 2.6\% (MS-MARCO), and 0.8\% (PubMedQA); Recall's MAD is approximately 30\% across all three --an improvement of $11\times$ to $37\times$. On the BEIR benchmarks, $1/\text{Ratio}$ achieves a MAD of 0.6\% (SciFact) and 0.5\% (NFCorpus), compared to Recall's 28.7\% --a $47\times$ to $57\times$ improvement. Across all five datasets, $1/\text{Ratio}$ is consistently a much better predictor of downstream RAG quality than Recall.

\subsection{Summary of Experimental findings}\label{sec:sumexp}

Our experiments support four main conclusions.
\textbf{Optimizing for $1/\text{Ratio}$ is substantially cheaper than optimizing for Recall (RQ1).}
Reaching a fixed quality threshold under $1/\text{Ratio}$ requires far less work than under Recall: at $T=0.95$ and $k=100$, satisfying Recall costs on average $1.86\times$-$9.36\times$ more distance computations (up to $33.3\times$ for HNSW on ImageNet), yields an average $3.27\times$ lower QPS, and takes up to $3.63\times$ (HNSW) and $2.82\times$ (SymphonyQG) more build time. The gap widens with $k$ and, in most cases, with intrinsic dimensionality, and Recall frequently rejects configurations that $1/\text{Ratio}$ accepts as distance-accurate.
\textbf{The advantage of $1/\text{Ratio}$ is computational rather than spatial (RQ1).}
The memory footprint of ANN methods is  %$1/\text{Ratio}$ is computational rather than spatial: search-phase memory is 
dominated by the base index structure rather than the parameters for achieving high Recall, so it is similar under both metrics.
\textbf{Algorithm rankings are mostly stable (RQ2).}
Replacing Recall with $1/\text{Ratio}$ does not change which algorithms are faster, lighter, or cheaper to build, only the absolute cost at which each reaches a given quality level. The relative magnitude of the differences; however, is metric-specific and can widen or narrow.
\textbf{Recall overstates the cost of approximation (RQ3, RQ4).}
The two metrics diverge as $k$ and LID grow: an algorithm increasingly returns neighbors that are as close as the true $k$NNs but carry different identifiers, penalizing Recall while leaving $1/\text{Ratio}$ high. Low Recall values have little effect on downstream task quality. Across image classification and RAG, retrieval quality stays stable as Recall falls to $0.4$, and $1/\text{Ratio}$ tracks true quality far more faithfully (prediction error stays under $3\%$, against $25$-$30\%$ for Recall).
\section{Conclusions}\label{sec:conclusions}

In this paper we questioned the long-standing convention of evaluating and tuning ANN algorithms based on Recall@$k$, and investigated 1/Ratio@$k$, the inverse approximation ratio, as a target quality measure. Unlike recently proposed alternatives, 1/Ratio@$k$ is simultaneously judge-free, hyperparameter-free, and computable from the embeddings and ground-truth identifiers that standard ANN benchmarks already provide. Benchmarking five state-of-the-art algorithms across six datasets of varying intrinsic dimensionality, we found that Recall@$k$ systematically overstates the cost of approximation: optimizing for 1/Ratio@$k$ reaches the same operational quality thresholds at substantially lower cost in query time, distance computations, and index build time, with the gap widening as $k$ and intrinsic dimensionality grow, while leaving the relative ranking of algorithms essentially unchanged. Crucially, index configurations that would be dismissed due to low Recall, 
%dismisses as inadequate, 
are not inaccurate in practice: across image classification and retrieval-augmented generation, downstream quality remained stable even as Recall dropped to 0.4, while 1/Ratio@$k$ tracked quality more faithfully.
% tracked retrieval quality far more faithfully than Recall did. 

The key takeouts from this study are (i) {\em by using 1/Ratio@$k$ as an objective instead of Recall@k,  researchers and practitioners can evaluate and tune equally effective ANN methods much faster}; and (ii) {\em expensive ANN index configurations that target high Recall levels can be replaced by faster and cheaper-to-construct configurations that achieve high 1/Ratio@$k$, without affecting the actual quality of the intended downstream tasks}.
%Taken together, these results suggest that the community has been tuning ANN methods toward a target that overstates the difficulty of the task, and that 1/Ratio@$k$ offers a more accurate and readily deployable proxy for real-world performance.

\bibliographystyle{ACM-Reference-Format}
\bibliography{citations}

\end{document}